\input amstex
\documentstyle{amsppt}
\def\const{\operatorname{const}}
\def\id{\operatorname{id}}
\def\sll{\operatorname{sl}}
\def\SL{\operatorname{SL}}
\topmatter
\title
On the point transformations for the equation
$y''=P+3\,Q\,y'+3\,R\,{y'}^2+S\,{y'}^3$.
\endtitle
\rightheadtext{On the point transformations \dots}
\author
R.A.~Sharipov
\endauthor
\thanks
This work was supported in part by European Fund INTAS (project
\#93-47, coordinator of project: S.I.~Pinchuk), by Russian Fund for
Fundamental Investigations (project \#96-01-00127, head of project:
Ya.T.~Sultanaev), and by the Academy of Sciences of the Republic
Bashkortostan (head of project N.M.~Asadullin).
\endthanks
\address
Bashkir State University, Frunze str. 32, 450074, Ufa, Russia
\endaddress
\email
root\@bgua.bashkiria.su
\endemail
\abstract For the equations $y''=P(x,y)+3\,Q(x,y)\,y'+3\,R(x,y)\,
{y'}^2+S(x,y)\,{y'}^3$ the problem of equivalence is considered.
Some classical results are resumed in order to prepare the
background for the study of special subclass of such equations,
which arises in the theory of dynamical systems admitting
the normal shift.
\endabstract
\endtopmatter
\document
\head
1. Introduction.
\endhead
     Let's consider an ordinary differential equation of the
second order with polynomial in $y'$ right hand side
of the following form:
$$
y''=P(x,y)+3\,Q(x,y)\,y'+3\,R(x,y)\,(y')^2+S(x,y)\,(y')^3.
\tag1.1
$$
Class of the equations \thetag{1.1} is invariant under the
point transformations
$$
\cases
\tilde x=\tilde x(x,y),\\
\tilde y=\tilde y(x,y).
\endcases
\tag1.2
$$
This means that after implementing the change of variables
\thetag{1.2} in such equation we obtain another equation
of the same form:
$$
\tilde y''=\tilde P(\tilde x,\tilde y)+3\,\tilde Q(\tilde x,
\tilde y)\,\tilde y'+3\,\tilde R(\tilde x,\tilde y)\,(\tilde y')^2
+\tilde S(\tilde x,\tilde y)\,(\tilde y')^3.
\tag1.3
$$
Suppose that two particular equations of the form
\thetag{1.1} are taken. The question on the existence of
the point transformation \thetag{1.2} that transfer one
of them into another is known as {the problem of equivalence}.
The study of this problem has the long history (see
\cite{1--22}). This paper is aimed to sum up all results
concerning the problem of equivalence for the equations
\thetag{1.1}, which are known to us, and prepare the background
for the study of special subclass of such equations:
$$
y''=\Phi_y(x,y)\,(y'+1)+\Phi_x(x,y).
$$
These equations arise in the theory of dynamical systems admitting
the normal shift (see paper \cite{23} for details).\par
\head
2. Point transformations and point symmetries.
\endhead
     Let's treat the pair of variables $(x,y)$ in \thetag{1.1}
as the coordinates of some point on a plane. Then the change
of variables \thetag{1.2} can be treated as the change of one
(curvilinear) system of coordinates for another. By such
treatment the equation \thetag{1.1} defines some geometrical
structure on the plane. On fixing some local coordinates this
structure is expressed by four functions $P(x,y)$, $Q(x,y)$,
$R(x,y)$\linebreak and $S(x,y)$.\par
     We shall take the point transformation \thetag{1.2} to
be regular. Denote by $S$ and $T$ direct and inverse transition
matrices for the point transformation \thetag{1.2};
$$
\xalignat 2
&S=\Vmatrix
x_{\sssize 1.0} &x_{\sssize 0.1}\\
\vspace{1ex}
y_{\sssize 1.0} &y_{\sssize 0.1}
\endVmatrix,
&&T=\Vmatrix
\tilde x_{\sssize 1.0} &\tilde x_{\sssize 0.1}\\
\vspace{1ex}
\tilde y_{\sssize 1.0} &\tilde y_{\sssize 0.1}
\endVmatrix.
\tag2.1
\endxalignat
$$
By means of double indices in \thetag{2.1} here and in what
follows we denote partial derivatives, e\. g\. for the function
$f(u,v)$ by $f_{\sssize p.q}$ we denote the derivative of
$p$-th order in the first argument $u$, and of $q$-th order in
the second argument $v$.
\par
     Formula for transforming the first derivatives by the point
transformation \thetag{1.2} has the following form:
$$
y'=\frac{y_{\sssize 1.0}+y_{\sssize 0.1}\,\tilde y'}
{x_{\sssize 1.0}+x_{\sssize 0.1}\,\tilde y'}.
\tag2.2
$$
Analogous formula is available for transforming the second order
derivatives:
$$
\aligned
y''&=\frac{(x_{\sssize 1.0}+x_{\sssize 0.1}\,\tilde y')
(y_{\sssize 2.0}+2\,y_{\sssize 1.1}\,\tilde y'+y_{\sssize 0.2}
\,(\tilde y')^2+y_{\sssize 0.1}\,\tilde y'')}
{(x_{\sssize 1.0}+x_{\sssize 0.1}\,\tilde y')^3}-\\
\vspace{2ex}
&\qquad-\frac{(y_{\sssize 1.0}+y_{\sssize 0.1}
\,\tilde y')(x_{\sssize 2.0}+2\,x_{\sssize 1.1}\,\tilde y'+
x_{\sssize 0.2}\,(\tilde y')^2+x_{\sssize 0.1}\,\tilde y'')}
{(x_{\sssize 1.0}+x_{\sssize 0.1}\,\tilde y')^3}.
\endaligned
\tag2.3
$$
By substituting \thetag{2.2} and \thetag{2.3} into \thetag{1.1},
we derive the transformation rules for the coefficients of the
equation \thetag{1.1}. In order to express these rules in compact
form, we introduce $3$-dimensional array with the following
components:
$$
\xalignat 2
&\theta_{111}=P,
&&\theta_{112}=\theta_{121}=\theta_{211}=Q,
\hskip -2em\\
\vspace{-1.7ex}
&&&\tag2.4\\
\vspace{-1.7ex}
&\theta_{122}=\theta_{212}=\theta_{221}=R,
&&\theta_{222}=S.
\hskip -2em
\endxalignat
$$
As we can see from \thetag{2.4}, the array $\theta_{ijk}$ is
symmetric with respect to any pair of indices. Let's raise
one of these indices
$$
\theta^k_{ij}=\sum^2_{r=1}d^{kr}\,\theta_{rij}
\tag2.5
$$
by means of skew-symmetric matrix $d^{ij}$ with the
following components:
$$
d_{ij}=d^{ij}=
\Vmatrix\format \r&\quad\l\\ 0 & 1\\-1 & 0\endVmatrix.
\tag2.6
$$
We are able to write the transformation rule for the quantities
$\theta^k_{ij}$ defined in \thetag{2.5}. By the change of variables
\thetag{1.2} they are transformed as follows:
$$
\theta^k_{ij}=\sum^2_{m=1}\sum^2_{p=1}\sum^2_{q=1}
S^k_m\,T^p_i\,T^q_j\,\tilde\theta^m_{pq}
+\sum^2_{m=1}S^k_m\,\frac{\partial T^m_i}{\partial x^j}-
\frac{\tilde\sigma_i\,\delta^k_j+
\tilde\sigma_j\,\delta^k_i}{3},
\tag2.7
$$
where $x^1=x$, $x^2=y$, $\tilde x^1=\tilde x$, $\tilde x^2=\tilde y$,
and where we used the following notations:
$$
\xalignat 3
&\qquad\tilde\sigma_i=\frac{\partial\ln\det T}{\partial x^i},
&&\delta^k_i=
\cases
1&\text{for\ }i=k,\\
0&\text{for\ }i\neq k.
\endcases
\tag2.8
\endxalignat
$$
Last summand with the quantities $\tilde\sigma_i$ and
$\tilde\sigma_j$ differs the formula \thetag{2.7} from the
standard transformation rule for the components of affine
connection (see \cite{24}). This is the very circumstance
that determines the complexity
of the structures connected with the equation \thetag{1.1}.
It prevent us from direct use of standard methods of
differential geometry.\par
    Apart from the treatment just described, one has another
treatment for the change of variables \thetag{1.2}. Taking
local coordinates on the plane for fixed, we can treat
\thetag{1.2} as a map $f$ which transform the point with
coordinates $(x,y)$ into the point with coordinates $(\tilde
x,\tilde y)$. Each solution of the equation \thetag{1.1} is
the function $y(x)$, its graph is some curve on the plane.
The transformation $f$ maps this curve onto another curve,
which also can be considered as the graph of some function.
\definition{Definition 2.1} The transformation of the plane
$f$ is called {\it the point symmetry} of the equation
\thetag{1.1} if it maps each solution of this equation into
another solution of the same equation.
\enddefinition
    The rules of transformation for the coefficients of the
equation \thetag{1.1} under the mapping of the plane $f$
are determined by the same formula \thetag{2.7}. However
the Jacobi matrices $T$ and $S$ are now interpreted as
differentials of direct and inverse maps $f_*$ and $f_*^{-1}$.
According to the definition~2.1, the map $f$ is the point
symmetry for the equation \thetag{1.1} if it preserves the
geometrical structure connected with this equation. If $f$
is the point symmetry for some equation \thetag{1.1}, then
$f^{-1}$ is also the point symmetry for the same equation.
Composition of two point symmetries of the given equation
is the point symmetry for this equation. Therefore the set
of point symmetries of the given equation forms the group,
which is the important characteristic of the geometrical
structure connected with this equation.\par
    Let's consider one-parametric subgroups in the group
of point symmetries of the equation \thetag{1.1}. In order
to do this we equip the map $f$ with additional numeric
parameter $t\in\Bbb R$. The property being a group make
some limitations for the dependence of $f_t$ on $t$:
$$
\xalignat 2
&f_0=\id,
&&f_{t+\tau}=f_t\,{\sssize\circ}\,f_\tau.
\tag2.9
\endxalignat
$$
Let's write the transformations \thetag{2.9} in coordinates
expanding them into the Tailor series in $t$ centered at the
point $t=0$:
$$
\aligned
&x\mapsto f^1_t(x,y)=x+t\cdot V(x,y)+\ldots,\\
&y\mapsto f^2_t(x,y)=y+t\cdot U(x,y)+\ldots.
\endaligned
\tag2.10
$$
The quantities $Z^1=V$ and $Z^2=U$ in the expansions \thetag{2.10}
appear to be the components of some vector field $\bold Z$ on the
plane. Knowing this field, we can recover the transformations $f_t$
if they form one-parametric group \thetag{2.9}. Derivatives $y'$
and $y''$ by the map \thetag{2.10} are transformed as follows:
$$
\xalignat 2
&y'\mapsto y'+t\cdot W_1+\ldots,
&&y''\mapsto y''+t\cdot W_2+\ldots.
\tag2.11
\endxalignat
$$
The quantities $W_1$ and $W_2$ are known as the
components of {\it the first extension} and {\it the
second extension} of the vector field $\bold Z$ (see
\cite{25} and \cite{26}). Their forms are determined by
the following formulas:
$$
\align
&W_1=U_{\sssize 1.0}+(U_{\sssize 0.1}-V_{\sssize 1.0})\,y'-
V_{\sssize 0.1}\,(y')^2,\tag2.12\\
\vspace{1.5ex}
&\aligned
 W_2&=(U_{\sssize 0.1}- 2\,V_{\sssize 1.0}-3\,V_{\sssize 0.1}\,y')
  \,y''+U_{\sssize 2.0}+\\
  \vspace{0.5ex}
  &+(2\,U_{\sssize 1.1}-V_{\sssize 2.0})\,y'+(U_{\sssize 0.2}
  -2\,V_{\sssize 1.1})\,(y')^2-V_{\sssize 0.2}\,(y')^3.
  \endaligned
  \tag2.13
\endalign
$$
Let's substitute \thetag{2.11} into the equation \thetag{1.1}
and take into account the relationships \thetag{2.12} and
\thetag{2.13}. Moreover, we take into account the following
expansions
$$
\xalignat 2
&P\mapsto P+(P_{\sssize 1.0}\,V+P_{\sssize 0.1}\,U)\,t+\ldots,
&&Q\mapsto Q+(Q_{\sssize 1.0}\,V+Q_{\sssize 0.1}\,U)\,t+\ldots,\\
\vspace{1.5ex}
&R\mapsto R+(R_{\sssize 1.0}\,V+R_{\sssize 0.1}\,U)\,t+\ldots,
&&S\mapsto S+(S_{\sssize 1.0}\,V+S_{\sssize 0.1}\,U)\,t+\ldots,
\endxalignat
$$
which are the consequences of \thetag{2.10}. Then the condition
of the existence of one-parametric group of point symmetries
for the equation \thetag{1.1} is expressed in form of the
system of four equations for two components of the vector
field $\bold Z$:
$$
\align
&\aligned
 U_{\sssize 2.0}&=3\,Q\,U_{\sssize 1.0}+P_{\sssize 0.1}\,U-
 P\,U_{\sssize 0.1}+2\,P\,V_{\sssize 1.0}+P_{\sssize 1.0}\,V,\\
 \vspace{1ex}
 -V_{\sssize 0.2}&=3\,R\,V_{\sssize 0.1}+S_{\sssize 1.0}\,V-
 S\,V_{\sssize 1.0}+2\,S\,U_{\sssize 0.1}+S_{\sssize 0.1}\,U,
 \endaligned\\
\vspace{-0.8ex}
& \tag2.14\\
\vspace{-3ex}
&\aligned
 2\,U_{\sssize 1.1}&=3\,Q\,V_{\sssize 1.0}+
 3\,Q_{\sssize 1.0}\,V+3\,Q_{\sssize 0.1}\,U+
 6\,R\,U_{\sssize 1.0}+3\,P\,V_{\sssize 0.1}+
 V_{\sssize 2.0},\\
 \vspace{1ex}
 -2\,V_{\sssize 1.1}&=3\,R\,U_{\sssize 0.1}
 +3\,R_{\sssize 0.1}\,U+3\,R_{\sssize 1.0}\,V
 +6\,Q\,V_{\sssize 0.1}+3\,S\,U_{\sssize 1.0}
 -U_{\sssize 0.2}.
 \endaligned
\endalign
$$
This system of four equations \thetag{2.14} is overdetermined.
It has a lot of differential consequences. First four of them
can be solved with respect to the higher order derivatives
$U_{\sssize 3.0}$, $U_{\sssize 2.1}$, $U_{\sssize 1.2}$, and
$U_{\sssize 0.3}$. They are the following equations
$$
\gather
\aligned
 2\,&U_{\sssize 3.0}=(2\,P_{\sssize 0.1}+18\,Q^2+6\,Q_{\sssize 1.0}-
 6\,P\,R)\,U_{\sssize 1.0}+\\
 &+(2\,P_{\sssize 2.0}+6\,Q\,P_{\sssize 1.0}-3\,P\,Q_{\sssize 1.0})\,V+
 3\,P\,V_{\sssize 2.0}+\\
 &+(2\,P_{\sssize 1.1}-3\,P\,Q_{\sssize 0.1}+6\,Q\,P_{\sssize 0.1})\,U
 -3\,P^2\,V_{\sssize 0.1}+\\
 &+(- 2\,P_{\sssize 1.0}-6\,Q\,P_{\sssize 0.0})\,U_{\sssize 0.1}+
 (9\,Q\,P+6\,P_{\sssize 1.0})\,V_{\sssize 1.0},
\endaligned\tag2.15\\
\allowdisplaybreak
\vspace{1ex}
\aligned
 2\,&U_{\sssize 2.1}=
 (2\,P_{\sssize 1.1}+9\,Q\,Q_{\sssize 1.0}-6\,P\,R_{\sssize 1.0})\,V+\\
 &+(- 6\,P\,R_{\sssize 0.1}+9\,Q\,Q_{\sssize 0.1}+2\,P_{\sssize 0.2})\,U+
 3\,Q\,V_{\sssize 2.0}+\\
 &+(-6\,P\,S+6\,Q_{\sssize 0.1}+18\,R\,Q)\,U_{\sssize 1.0}-
 6\,R\,P\,U_{\sssize 0.1}+\\
 &+(9\,Q^2+4\,P_{\sssize 0.1})\,V_{\sssize 1.0}+
 (- 3\,Q\,P+2\,P_{\sssize 1.0})\,V_{\sssize 0.1},
\endaligned\tag2.16\\
\allowdisplaybreak
\vspace{1ex}
\aligned
 2\,&U_{\sssize 1.2}=
 (9\,R\,Q_{\sssize 1.0}+4\,Q_{\sssize 1.1}-2\,R_{\sssize 2.0}
 -2\,S\,P_{\sssize 1.0}-4\,P\,S_{\sssize 1.0})\,V+\\
 &+(4\,Q_{\sssize 0.2}-2\,R_{\sssize 1.1}-2\,S\,P_{\sssize
0.1}+9\,R\,Q_{\sssize 0.1}-4\,P\,S_{\sssize 0.1})\,U+
 3\,R\,V_{\sssize 2.0}+\\
 &+(6\,R_{\sssize 0.1}-6\,S\,Q-2\,S_{\sssize 1.0}+
 18\,R^2)\,U_{\sssize 1.0}+
 (4\,P_{\sssize 0.1}-3\,P\,R)\,V_{\sssize 0.1}+\\
 &+(4\,Q_{\sssize 0.1}-6\,P\,S-2\,R_{\sssize 1.0})\,U_{\sssize 0.1}+
 (9\,R\,Q+4\,Q_{\sssize 0.1}-2\,R_{\sssize 1.0})\,V_{\sssize 1.0},
\endaligned\tag2.17\\
\allowdisplaybreak
\vspace{1ex}
\aligned
 2\,&U_{\sssize 0.3}=
 (18\,R\,R_{\sssize 1.0}-3\,S\,Q_{\sssize 1.0}+6\,R_{\sssize 1.1}-
 4\,S_{\sssize 2.0}-12\,Q\,S_{\sssize 1.0})\,V+\\
 &+(6\,R_{\sssize 0.2}-3\,S\,Q_{\sssize 0.1}-12\,Q\,S_{\sssize 0.1}-
 4\,S_{\sssize 1.1}+18\,R\,R_{\sssize 0.1})\,U+\\
 &+(2\,S_{\sssize 0.1}+12\,R\,S)\,U_{\sssize 1.0}+(12\,Q_{\sssize 0.1}-
 6\,R_{\sssize 1.0}-3\,P\,S_{\sssize 0.0})\,V_{\sssize 0.1}+\\
 &+(12\,R_{\sssize 0.1}+18\,R^2-8\,S_{\sssize 1.0}-
 24\,S\,Q)\,U_{\sssize 0.1}+9\,Q\,S\,V_{\sssize 1.0}+
 3\,S\,V_{\sssize 2.0}.\\
\endaligned\tag2.18
\endgather
$$
In addition to \thetag{2.15}, \thetag{2.16}, \thetag{2.17}, and
\thetag{2.18}, the equations \thetag{2.14} has next four
differential consequences with derivatives $V_{\sssize 3.0}$,
$V_{\sssize 2.1}$, $V_{\sssize 1.2}$, and $V_{\sssize 0.3}$:
$$
\gather
\aligned
 2\,&V_{\sssize 0.3}=
 (18\,R^2-6\,S\,Q-6\,R_{\sssize 0.1}-2\,S_{\sssize 1.0})
 \,V_{\sssize 0.1}+\\
 &+(6\,R\,S_{\sssize 0.1} - 3\,S\,R_{\sssize 0.1}
 - 2\,S_{\sssize 0.2})\,U
 -3\,S\,U_{\sssize 0.2}+\\
 &+(6\,R\,S_{\sssize 1.0}-2\,S_{\sssize 1.1}-3\,S\,R_{\sssize 1.0})\,V-
 3\,S^2\,U_{\sssize 1.0}+\\
 &+(9\,R\,S-6\,S_{\sssize 0.1})\,U_{\sssize 0.1}+
 (- 6\,R\,S+2\,S_{\sssize 0.1})\,V_{\sssize 1.0},
\endaligned\tag2.19\\
\allowdisplaybreak
\vspace{1ex}
\aligned
 2\,&V_{\sssize 1.2}=
 (9\,R\,R_{\sssize 0.1}-6\,S\,Q_{\sssize 0.1}-2\,S_{\sssize 1.1})\,U+\\
 &+(9\,R\,R_{\sssize 1.0}-6\,S\,Q_{\sssize 1.0}-2\,S_{\sssize 2.0})\,V
 -3\,R\,U_{\sssize 0.2}+\\
 &+(18\,R\,Q-6\,P\,S-6\,R_{\sssize 1.0})\,V_{\sssize 0.1}
 -6\,Q\,S\,V_{\sssize 1.0}-\\
 &-(3\,R\,S+2\,S_{\sssize 0.1})\,U_{\sssize 1.0}
 +(9\,R^{2}-4\,S_{\sssize 1.0})\,U_{\sssize 0.1},
\endaligned\tag2.20\\
\allowdisplaybreak
\vspace{1ex}
\aligned
 2\,&V_{\sssize 2.1}=
 (2\,Q_{\sssize 0.2}-4\,S\,P_{\sssize 0.1}-4\,R_{\sssize 1.1}
 -2\,P\,S_{\sssize 0.1}+9\,Q\,R_{\sssize 0.1})\,U+\\
 &+(9\,Q\,R_{\sssize 1.0}-2\,P\,S_{\sssize 1.0}+2\,Q_{\sssize 1.1}
 -4\,R_{\sssize 2.0}-4\,S\,P_{\sssize 1.0})\,V-3\,Q\,U_{\sssize 0.2}+\\
 &+(2\,Q_{\sssize 0.1}+9\,R\,Q-4\,R_{\sssize 1.0})\,U_{\sssize 0.1}
 +(2\,Q_{\sssize 0.1}-4\,R_{\sssize 1.0}-6\,P\,S)\,V_{\sssize 1.0}+\\
 &+(18\,Q^2-6\,Q_{\sssize 1.0}
 -6\,P\,R+2\,P_{\sssize 0,1})\,V_{\sssize 0.1}
 -(4\,S_{\sssize 1.0}+3\,S\,Q)\,U_{\sssize 1.0},
\endaligned\tag2.21\\
\allowdisplaybreak
\vspace{1ex}
\aligned
 2\,&V_{\sssize 3.0}=
 (4\,P_{\sssize 0.2}-3\,P\,R_{\sssize 0.1}-12\,R\,P_{\sssize 0.1}
 -6\,Q_{\sssize 1.1}+18\,Q\,Q_{\sssize 0.1})\,U+\\
 &+(18\,Q\,Q_{\sssize 1.0}+4\,P_{\sssize 1.1}-12\,R\,P_{\sssize 1.0}
 -6\,Q_{\sssize 2.0}-3\,P\,R_{\sssize 1.0})\,V+\\
 &+(6\,Q_{\sssize 0.1}-12\,R_{\sssize 1.0}-3\,P\,S)\,U_{\sssize 1.0}
 +(12\,Q\,P-2\,P_{\sssize 1.0})\,V_{\sssize 0.1}+\\
 &+(18\,Q^{2}-12\,Q_{\sssize 1.0}-24\,P\,R
 +8\,P_{\sssize 0.1})\,V_{\sssize 1.0}+9\,R\,P\,U_{\sssize 0.1}
 -3\,P\,U_{\sssize 0.2}.
\endaligned\tag2.22
\endgather
$$
Right hand sides of the equations \thetag{2.14},
\thetag{2.15}--\thetag{2.18} and the last four equations
\thetag{2.19}, \thetag{2.20}, \thetag{2.21}, \thetag{2.22}
contain the following functions: $V$, $V_{\sssize 1.0}$,
$V_{\sssize 0.1}$, $V_{\sssize 2.0}$, $U$, $U_{\sssize 0.1}$,
$U_{\sssize 1.0}$, $U_{\sssize 0.2}$. Lets form the vector-column
$\Psi$ with them. Then, using part of the above equations, we can
obtain the complete system of Pfaff equations with respect to the
components of the vector-column $\Psi$:
$$
\xalignat 2
&\partial_x\Psi=\Lambda_x\,\Psi,
&
&\partial_y\Psi=\Lambda_y\,\Psi.
\tag2.23
\endxalignat
$$
Here $\Lambda_x$ and $\Lambda_y$ are two square matrices
$8\times 8$. \pagebreak The compatibility condition for the
system of linear equations \thetag{2.23} is expressed as
a condition of commutating for two differential operators
with matrix coefficients:
$$
[\partial_x-\Lambda_x,\,\partial_y-\Lambda_y]=0,
\tag2.24
$$
In the theory of integrability the equation \thetag{2.24} is
known as {\it the equation of zero curvature}. Writing down
it in components we find that it is reduced to
$$
\xalignat 2
&A=0,
&
&B=0,
\tag2.25
\endxalignat
$$
where by $A$ and $B$ we denote the following quantities:
$$
\aligned
&\aligned
 A=P_{\sssize 0.2}&-2\,Q_{\sssize 1.1}+R_{\sssize 2.0}+
 2\,P\,S_{\sssize 1.0}+S\,P_{\sssize 1.0}-\\
 \vspace{0.5ex}
 &-3\,P\,R_{\sssize 0.1}-3\,R\,P_{\sssize 0.1}-
 3\,Q\,R_{\sssize 1.0}+6\,Q\,Q_{\sssize 0.1},
 \endaligned\\
 \vspace{1ex}
&\aligned
 B=S_{\sssize 2.0}&-2\,R_{\sssize 1.1}+Q_{\sssize 0.2}-
 2\,S\,P_{\sssize 0.1}-P\,S_{\sssize 0.1}+\\
 \vspace{0.5ex}
 &+3\,S\,Q_{\sssize 1.0}+3\,Q\,S_{\sssize 1.0}+
 3\,R\,Q_{\sssize 0.1}-6\,R\,R_{\sssize 1.0}.
 \endaligned
\endaligned
\tag2.26
$$
The quantities $\alpha_1=A$ and $\alpha_2=B$ appear to be the
components of pseudocovectorial field of the weight $1$. The
quantities
$$
\alpha^i=\sum^2_{k=1}d^{ik}\,\alpha_k
\tag2.27
$$
obtained from $\alpha_1=A$ and $\alpha_2=B$ by raising the
lower index are the components of pseudovectorial field of
the weight $2$. The definition of pseudotensorial field
just below can be found in \cite{27}.\par
\definition{Definition 2.2} Pseudotensorial field of the type
$(r,s)$ and weight $m$ is an array of quantities $F^{i_1\ldots\,
i_r}_{j_1\ldots\,j_s}$ which under the change of variables
\thetag{1.2} transforms as
$$
F^{i_1\ldots\,i_r}_{j_1\ldots\,j_s}=
(\det T)^m\sum\Sb p_1\ldots p_r\\ q_1\ldots q_s\endSb
S^{i_1}_{p_1}\ldots\,S^{i_r}_{p_r}\,\,
T^{q_1}_{j_1}\ldots\,T^{q_s}_{j_s}\,\,
\tilde F^{p_1\ldots\,p_r}_{q_1\ldots\,q_s}.
\tag2.28
$$
\enddefinition
     Check of the rule \thetag{2.28} for the quantities
$\alpha_k$ and for the quantities $\alpha^i$ in \thetag{2.27}
can be done by direct calculations based on \thetag{2.7}.
The following lemma is a consequence of pseudotensorial
character of the transformation rule for $A$ and $B$.
\proclaim{Lemma 2.1} Simultaneous vanishing of the parameters
$A=0$ and $B=0$ for the initial equation \thetag{1.1} is
equivalent to their vanishing $\tilde A=0$ and $\tilde B=0$
for the transformed equation \thetag{1.3}.
\endproclaim\pagebreak
     Vanishing conditions \thetag{2.25} determine {the case
of maximal degeneration} in the theory of the equations
\thetag{1.1}. This case is considered in the next section.\par
\head
3. Case of maximal degeneration.
\endhead
     Let's take the conditions \thetag{2.25} to be fulfilled.
These conditions are invariant with respect to the point
transformations (see lemma~2.1 above). As a consequence of
\thetag{2.25} we get the compatibility of the Pfaff equations
\thetag{2.23}. It is known that the set of solutions of complete
compatible system of Pfaff equations is a finite dimensional
linear vector-space over the real numbers $\Bbb R$. Its dimension
coincides with the number of component of the vector-column $\Psi$,
which is equal to eight.\par
     The equations \thetag{2.14} are contained in the system of
equations \thetag{2.23}, therefore each solution of this system
determines the vector field $\bold Z$ for some one-parametric
group of point symmetries of the equation \thetag{1.1}. It is
known that the set of point symmetries of any equation is
closed with respect to the mutual commutators of corresponding
vector fields. Therefore the vector fields of point symmetries
of the equation \thetag{1.1} form $8$-dimensional Lie algebra
over real numbers, provided the conditions \thetag{2.25} are
fulfilled. Components of vector-column $\Psi$ in some fixed
point (e. g. the point $x=0$ and $y=0$) are the natural coordinates
in this algebra. We shall not calculate the structural constants of
this algebra now, but instead of this we shall prove the following
theorem.
\proclaim{Theorem 3.1} The equation \thetag{1.1} can be
transformed to the form $\tilde y''=0$ by means of some point
transformation if and only if the relationships \thetag{2.25}
are fulfilled.
\endproclaim
\demo{Proof} Suppose that the equation \thetag{1.1} is reducible
to the equation $\tilde y''=0$. The conditions $\tilde P=0$,
$\tilde Q=0$, $\tilde R=0$, and $\tilde S=0$ for the coefficients
of the transformed equation are written as the four differential
equations for two functions $\tilde x(x,y)$ and $\tilde y(x,y)$
determining the appropriate point transformation \thetag{1.2}
They are derived from \thetag{2.7}. The equation $\tilde P=0$
has the form:
$$
(\tilde x_{\sssize 2.0}
+P\,\tilde x_{\sssize 0.1})\,\tilde y_{\sssize 1.0}=
(\tilde y_{\sssize 2.0}
-P\,\tilde y_{\sssize 0.1})\,\tilde x_{\sssize 1.0}.
\tag3.1
$$
For the regular transformation \thetag{1.2} at least one of the
derivatives $\tilde x_{\sssize1.0}$ or $\tilde y_{\sssize1.0}$
in \thetag{3.1} is nonzero. Therefore, after introducing an
additional function $M(x,y)$, we can replace \thetag{3.1} by
two equations:
$$
\xalignat 2
&\tilde x_{\sssize 2.0}=M\,\tilde x_{\sssize 1.0}
-P\,\tilde x_{\sssize 0.1},
&
&\tilde y_{\sssize 2.0}=M\,\tilde y_{\sssize 1.0}
-P\,yx_{\sssize 0.1}.
\tag3.2
\endxalignat
$$
In the similar way, the equation $\tilde S=0$ can be broken
into two equations after introducing another additional
function $N(x,y)$:
$$
\xalignat 2
&\tilde x_{\sssize 0.2}=S\,\tilde x_{\sssize 1.0}
-N\,\tilde x_{\sssize 0.1},
&
&\tilde y_{\sssize 0.2}=S\,\tilde y_{\sssize 1.0}
-N\,yx_{\sssize 0.1}.
\tag3.3
\endxalignat
$$
Then the rest two equations $\tilde Q=0$ and $\tilde R=0$
\pagebreak can be used to determine the partial derivatives
$\tilde x_{\sssize 1.1}$ and $\tilde y_{\sssize 1.1}$. For
these two partial derivatives we obtain:
$$
\aligned
&\tilde x_{\sssize 1.1}=\frac{3\,R-N}{2}\,\tilde x_{\sssize 1.0}
-\frac{3\,Q-N}{2}\,\tilde x_{\sssize 0.1},\\
\vspace{2ex}
&\tilde y_{\sssize 1.1}=\frac{3\,R-N}{2}\,\tilde y_{\sssize 1.0}
-\frac{3\,Q-N}{2}\,\tilde y_{\sssize 0.1}.
\endaligned
\tag3.4
$$
As a result, for the function $\tilde x(x,y)$ we get the system
of three differential equations, which can be expressed as
the complete system of Pfaff equations \thetag{2.23} with
respect to the vector column $\Psi$ composed by two derivatives
$\tilde x_{\sssize 1.0}$ and $\tilde x_{\sssize 0.1}$. The
equations for the function $\tilde y(x,y)$ have exactly the
same form.\par
     By introducing the additional functions $M$ and $N$
we managed to separate dependent variables $\tilde x$ and
$\tilde y$ in the equations $\tilde P=0$, $\tilde Q=0$,
$\tilde R=0$, $\tilde S=0$, replacing them by two
identical systems of equations. Being absolutely identical
these two systems of equations have common compatibility
condition, which is written as the matrix equation
\thetag{2.24}. In the present case this matrix equation
is equivalent to the system of four equations for the
functions $M$ and $N$:
$$
\gather
\aligned
&M_{\sssize 1.0}=3\,Q_{\sssize 1.0}-2\,P_{\sssize 0.1}
+\frac{M^2}{2}+P\,N+3\,P\,R-\frac{9\,Q^2}{2},\\
\vspace{2ex}
&N_{\sssize 1.0}=2\,Q_{\sssize 0.1}-R_{\sssize 1.0}
-\frac{3\,M\,R}{2}-2\,P\,S+\frac{M\,N}{2}
-\frac{3\,Q\,N}{2}+\frac{9\,Q\,R}{2},\hskip -3em
\endaligned
\tag3.5\\
\vspace{2ex}
\aligned
&M_{\sssize 0.1}=2\,R_{\sssize 1.0}-Q_{\sssize 0.1}
+\frac{3\,Q\,N}{2}+2\,S\,P-\frac{N\,M}{2}
+\frac{3\,R\,M}{2}-\frac{9\,R\,Q}{2},\hskip -3em\\
\vspace{2ex}
&N_{\sssize 0.1}=3\,R_{\sssize 0.1}-2\,S_{\sssize 1.0}
-\frac{N^2}{2}-S\,M-3\,S\,Q+\frac{9\,R^2}{2}.
\endaligned
\tag3.6
\endgather
$$
The equations \thetag{3.5} and \thetag{3.6} form one more
complete system of Pfaff equations. However this is the
system of nonlinear equations. Theory of nonlinear Pfaff
equations is given in the Appendix~A (see section~13 below).
By direct calculations based on this theory we find
that compatibility condition for the equations \thetag{3.5}
and \thetag{3.6} is exactly the condition \thetag{2.25}.
Thus we proved that condition \thetag{2.25} is necessary
for the equation \thetag{1.1} can be reduced to the form
$\tilde y''=0$.\par
     In order to prove sufficiency of this condition, we simply
do the above steps in the reverse order. From \thetag{2.25} we
obtain the compatibility of the equations \thetag{3.5} and
\thetag{3.6}. By solving them we find $M(x,y)$ and $N(x,y)$,
then substitute them into the equations \thetag{3.2},
\thetag{3.3}, and \thetag{3.4}. This make them compatible.
By solving the following initial value problem for these equations
$$
\pagebreak
\xalignat 2
&\tilde x_{\sssize 1.0}\,{\vrule height 8pt depth 14pt}
 \lower 7pt
 \hbox{\vtop{\baselineskip=0pt
             \lineskip=2pt
             \lineskiplimit=0pt
             \hbox{$\,\ssize x=a$}
             \hbox{$\,\ssize y=b$}}}
             =1,
&&\tilde x_{\sssize 0.1}\,{\vrule height 8pt depth 14pt}
 \lower 7pt
 \hbox{\vtop{\baselineskip=0pt
             \lineskip=2pt
             \lineskiplimit=0pt
             \hbox{$\,\ssize x=a$}
             \hbox{$\,\ssize y=b$}}}
             =0,\tag3.7\\
\allowdisplaybreak
\vspace{2ex}
&\tilde y_{\sssize 1.0}\,{\vrule height 8pt depth 14pt}
 \lower 7pt
 \hbox{\vtop{\baselineskip=0pt
             \lineskip=2pt
             \lineskiplimit=0pt
             \hbox{$\,\ssize x=a$}
             \hbox{$\,\ssize y=b$}}}
             =0,
&&\tilde y_{\sssize 0.1}\,{\vrule height 8pt depth 14pt}
 \lower 7pt
 \hbox{\vtop{\baselineskip=0pt
             \lineskip=2pt
             \lineskiplimit=0pt
             \hbox{$\,\ssize x=a$}
             \hbox{$\,\ssize y=b$}}}
             =1.\tag3.8
\endxalignat
$$
we get the pair of functions $\tilde x(x,y)$ and $\tilde y(x,y)$.
Due to initial data \thetag{3.7} and \thetag{3.8}, we have the
regularity of the point transformations \thetag{1.2} given by
these functions in some neighborhood of the point $x=a$ and
$y=b$. And finally, due to the equations \thetag{3.5} and
\thetag{3.6}, the transformed equation \thetag{1.3} has the
form $\tilde y''=0$.\qed\enddemo
     The theorem~3.1 just proved reduces all equations
\thetag{1.1} satisfying the condition \thetag{2.25} to the
trivial equation $y''=0$. Therefore we are to describe the
algebra of point symmetries only for this trivial equation.
Each element of this algebra can be represented in three
forms: in form of one-parametric group of point transformations,
in form of corresponding vector field with components $(V,U)$,
and in form of $8$-dimensional constant vector
$$
\Psi=(V,V_{\sssize 1.0},V_{\sssize 0.1},V_{\sssize 2.0},
U,U_{\sssize 0.1},U_{\sssize 1.0},U_{\sssize 0.2})
$$
composed of the values of the above functions at some fixed
point, e\. g\. at the point $x=0$ and $y=0$. First we
describe the base elements of the algebra of point symmetries
for the equation $y''=0$.\par
     1. {\it Shift of argument}: $x\mapsto x+t$, $y\mapsto y$.
Components of corresponding vector field are: $V=1$, $U=0$; and
$\Psi=(1,0,0,0,0,0,0,0)$.\par
    2. {\it Shift of function value}: $x\mapsto  x$,
$y\mapsto y+t$. Components of corresponding vector field are:
$V=0$, $U=1$; and $\Psi=(0,0,0,0,1,0,0,0)$.\par
    3. {\it Blowing up the argument}: $x\mapsto x\,e^t$,
$y\mapsto y$. Components of corresponding vector field are:
$V=x$, $U=0$; and $\Psi=(0,1,0,0,0,0,0,0)$.\par
    4. {\it Blowing up the function value}: $x\mapsto x$,
$y\mapsto y\,e^t$. Components of corresponding vector field are:
$V=0$, $U=y$; and $\Psi=(0,0,0,0,0,1,0,0)$.\par
    5. {\it Shift of argument with zoom}: $x\mapsto x+y\,t$,
$y\mapsto y$. Components of corresponding vector field are:
$V=y$, $U=0$; and $\Psi=(0,0,1,0,0,0,0,0)$.\par
    6. {\it Shift of function value with zoom}: $x\mapsto  x$,
$y\mapsto y+x\,t$. Components of corresponding vector field are:
$V=0$, $U=x$; and $\Psi=(0,0,0,0,0,0,1,0)$.\par
    7. {\it Inversion of the argument}. Components of corresponding
vector field are: $V=x^2/2$, $U=x\,y/2$; and $\Psi=(0,0,0,1,0,0,0,0)$.
One-parametric group of point transformations is given by
$$
\xalignat 2
&x\mapsto\frac{2\,x}{2-x\,t},
&&y\mapsto\frac{2\,y}{2-x\,t}.
\endxalignat
$$\par
    8. {\it Inversion of the function value}. Components of corresponding
vector field are: $V=x\,y/2$, $U=y^2/2$; and $\Psi=(0,0,0,0,0,0,0,1)$.
One-parametric group of point transformations is given by
$$
\xalignat 2
&x\mapsto\frac{2\,x}{2-y\,t},
&&y\mapsto\frac{2\,y}{2-y\,t}.
\endxalignat
$$\par
\proclaim{Theorem 3.2} Algebra of point symmetries of the equation
\thetag{1.1}, satisfying the condition \thetag{2.25}, is isomorphic
to the special linear Lie algebra $\sll(3,\Bbb R)$.
\endproclaim
\demo{Proof} Because of the theorem~3.1 we should prove this
statement only for the equation $y''=0$. In order to do this,
let's consider the map
$$
\Vmatrix
\psi_1\\ \psi_2\\ \psi_3\\ \psi_4\\
\psi_5\\ \psi_6\\ \psi_7\\ \psi_8
\endVmatrix
\mapsto
\Vmatrix
\dsize \frac{2\,\psi_2-\psi_6}{3} & \psi_3 & \psi_1\\
\vspace{3ex}
\psi_7 & \dsize\frac{2\,\psi_6-\psi_2}{3} & \psi_5 \\
\vspace{3ex}
\dsize -\frac{\psi_4}{2} & \dsize -\frac{\psi_8}{2} &
\dsize -\frac{\psi_2+\psi_6}{3}
\endVmatrix.
\tag3.9
$$
Now by means of direct calculations one can show that
\thetag{3.9} is the very map that implements the isomorphism
of the algebra of point symmetries of the equation $y''=0$ and
the matrix algebra $\sll(3,\Bbb R)$.
\qed\enddemo
    In terms of matrices from the group $\SL(3,\Bbb R)$ we can
describe all point transformations \thetag{1.2} that transform
the equation $y''=0$ into the same equation $\tilde y''=0$:
$$
\xalignat 2
&\tilde x=\frac{g^1_1\,x+g^1_2\,y+g^1_3}{g^3_1\,x+
g^3_2\,y+g^3_3},
&&\tilde y=\frac{g^2_1\,x+g^2_2\,y+g^2_3}{g^3_1\,x+
g^3_2\,y+g^3_3}.
\tag3.10
\endxalignat
$$
Formulas \thetag{3.10} can be derived by solving the equations
\thetag{3.5} and \thetag{3.6} for the case $P=Q=R=S=0$. Then
the obtained functions $M$ and $N$ should be substituted into
the equations \thetag{3.2}--\thetag{3.4}. Solutions of the
latter ones are the functions $\tilde x(x,y)$ and $\tilde y(x,y)$
of the form \thetag{3.10}.\par
\head
4. Case of general position.
\endhead
    Now suppose that parameters $A$ and $B$ in \thetag{2.26}
do not vanish simultaneously. Therefore we have two extra
compatibility equations for the components of the vector
field of the point symmetry
$$
\aligned
2\,V_{\sssize 1.0}\,A+V\,A_{\sssize 1.0}+U_{\sssize 1.0}\,B
+U_{\sssize 0.1}\,A+U\,A_{\sssize 0.1}=0,\\
\vspace{1ex}
V\,B_{\sssize 1.0}+2\,U_{\sssize 0.1}\,B+U\,B_{\sssize 0.1}
+V_{\sssize 0.1}\,A+V_{\sssize 1.0}\,B=0.
\endaligned
\tag4.1
$$
They appear as the following compatibility conditions for the
third order derivatives from \thetag{2.15}, \thetag{2.16},
\thetag{2.20}, and \thetag{2.22}:
$$
\xalignat 2
&\frac{\partial U_{\sssize 3.0}}{\partial y}=
\frac{\partial U_{\sssize 2.1}}{\partial x},
&&\frac{\partial V_{\sssize 0.3}}{\partial x}=
\frac{\partial V_{\sssize 1.2}}{\partial y}.
\endxalignat
$$
Other equations of this sort add nothing new to \thetag{4.1}.
\par
    The equations \thetag{4.1} are not unconditionally solvable
respective to any one of derivatives $U_{\sssize 1.0}$,
$U_{\sssize 0.1}$, $V_{\sssize 1.0}$, and $V_{\sssize 0.1}$ in
them. But we can overcome this difficulty. Let's remember that
the quantities $\alpha^1=B$ and $\alpha^2=-A$ are the components
of the pseudovectorial field of the weight $2$. This means that
we can find local coordinates on the plane such that $A=0$ and
$B\neq 0$. In order to do this, we should choose some vector field
collinear to $\alpha$ and then straighten it by changing for the
appropriate local coordinates.\par
     Now let's do one more point transformation that preserve
the condition $A=0$. It should be of the following form
$$
\xalignat 2
&\tilde x=\tilde x(x,y), &&\tilde y=y.
\tag4.2
\endxalignat
$$
The transformation rule for the pseudovector $\alpha$ under
the point transformation \thetag{4.2} can be written in
matrix form:
$$
\Vmatrix \tilde B\\ \vspace{2ex} 0\endVmatrix=
(\det T)^2\,
\Vmatrix x_{\sssize 1.0} & x_{\sssize 0.1}\\
\vspace{2ex}
0 & 1\endVmatrix\cdot\Vmatrix B\\ \vspace{2ex}
0\endVmatrix,
\tag4.3
$$
(see definition~2.2 and the formulas \thetag{2.1} and
\thetag{2.28}). From \thetag{4.3} we obtain\linebreak
$\tilde B=\tilde x_{\sssize 1.0}\,B$. By means of proper choice
of the function $\tilde x(x,y)$ in \thetag{4.2} we can
get $\tilde B=1$. Therefore we may take the following
condition
$$
\xalignat 2
&\alpha^2=A=0, &&\alpha^1=B=1
\tag4.4
\endxalignat
$$
to be true from the very beginning. Now substituting \thetag{4.4}
into \thetag{4.1} we can solve these equations with respect to
the following derivatives of the first order:
$$
\xalignat 2
&U_{\sssize 1.0}=0,&&V_{\sssize 1.0}=-2\,U_{\sssize 0.1}.
\tag4.5
\endxalignat
$$
From \thetag{4.5} we derive new compatibility conditions
$$
\xalignat 2
&U{\sssize 2.0}=\frac{\partial U_{\sssize 1.0}}{\partial x},
&&U{\sssize 1.1}=\frac{\partial U_{\sssize 1.0}}{\partial y}.
\tag4.6
\endxalignat
$$
Taking into account the equations \thetag{2.14}, from the first
equation \thetag{4.6} we get
$$
P\,U_{\sssize 0.1}=\frac{1}{5}\,P_{\sssize 1.0}\,V+
\frac{1}{5}\,P_{\sssize 0.1}\,U.
\tag4.7
$$
The second equation \thetag{4.6} can be solved with respect to
the derivative $V_{\sssize 2.0}$:
$$
V_{\sssize 2.0}=6\,Q\,U_{\sssize 0.1}-3\,Q_{\sssize 1.0}\,V
-3\,Q_{\sssize 0.1}\,U-3\,P\,V_{\sssize 0.1}.
\tag4.8
$$
Now, taking into account \thetag{4.8} and \thetag{2.14}, from
\thetag{4.5} we can derive next two relationships, which are
the compatibility conditions like \thetag{4.6}:
$$
\xalignat 2
&V{\sssize 2.0}=\frac{\partial V_{\sssize 1.0}}{\partial x},
&&V{\sssize 1.1}=\frac{\partial V_{\sssize 1.0}}{\partial y}.
\tag4.9
\endxalignat
$$
In explicit form, first of the compatibility equations \thetag{4.9}
is written as follows:
$$
P\,V_{\sssize 0.1}=2\,Q\,U_{\sssize 0.1}-Q_{\sssize 1.0}\,V
-Q_{\sssize 0.1}\,U,
\tag4.10
$$
Second compatibility equation \thetag{4.9} can be solved with
respect to the derivative of the second order $U_{\sssize 0.2}$:
$$
U_{\sssize 0.2}=\frac{6}{5}\,Q\,V_{\sssize 0.1}
+\frac{3}{5}\,R\,U_{\sssize 0.1}
+\frac{3}{5}\,R_{\sssize 1.0}\,V
+\frac{3}{5}\,R_{\sssize 0.1}\,U.
\tag4.11
$$
Further study of determining equations for the vector field
of point symmetry for the equation \thetag{1.1} is subdivided
into several cases. {\it The case of  general position} is
distinguished by the condition $P\neq 0$.\par
     Local coordinates satisfying the conditions \thetag{4.4}
are not determined uniquely: change from some particular
coordinates with this property to another ones is given by
the following relationships
$$
\xalignat 2
&\tilde x=\frac{x}{h'(y)^2}+g(y),
&&\tilde y=h(y),
\tag4.12
\endxalignat
$$
where $g(y)$ and $h(y)$ are two arbitrary functions of one
variable, and where $h'(y)\neq 0$.
Now let's write down the transition matrices for the
change of variables \thetag{4.12}:
$$
\xalignat 2
&T=\Vmatrix
\dsize\frac{1}{{h'}^2} &\dsize-\frac{2\,x\,h''}{{h'}^3}+g'\\
\vspace{4ex}
0 & h'
\endVmatrix.
&&S=\Vmatrix
{h'}^2 &\dsize\frac{2\,x\,h''}{{h'}^2}-g'\,h'\\
\vspace{2ex}
0 & \dsize\frac{1}{h'}
\endVmatrix.\hskip-2em
\tag4.13
\endxalignat
$$
Transformation rules for the coefficients of the equation
\thetag{1.1} are defined by \thetag{2.7}. However, due to
the special form of point transformation \thetag{4.12}, and
due to the special form of transition matrices \thetag{4.13},
these transformation rules are substantially simplified. The
most simple is the rule for coefficient $P$:
$$
P=\frac{\tilde P}{{h'}^5}=(\det T)^5\,\tilde P.
\tag4.14
$$
The transformation rule \thetag{4.14} let us define the
pseudoscalar field $F$ of the unit weight $1$. In order
to do it we set
$$
F^5=-P.
\tag4.15
$$
in some special local coordinates, where the conditions
\thetag{4.4} are fulfilled.\par
     Apart from the pseudoscalar field $F$ from \thetag{4.15}, we
introduce four quantities $\beta_1$, $\beta_2$, $\beta^1$, and
$\beta^2$. Let's define them by formulas
$$
\xalignat 2
&\beta_1=3\,P,&&\beta_2=3\,Q,
\tag4.16\\
\vspace{1ex}
&\beta^1=3\,Q,&&\beta^2=-3\,P
\tag4.17
\endxalignat
$$
in the above special coordinates. One can easily check that
by the transformations \thetag{4.12} the quantities \thetag{4.16}
and \thetag{4.17} are transformed as pseudocovectorial field of
the weight $3$ and as pseudovectorial field of the weight $4$
respectively.\par
     Using the pseudoscalar field $F$, we define the quantities
$\varphi_i$ as logarithmical derivatives of $F$:
$$
\varphi_i=-\frac{\partial\ln F}{\partial x^i}.
\tag4.18
$$
By the point transformations \thetag{4.12} these quantities are
transformed according to the following rule:
$$
\varphi_i=\sum^2_{j=1}T^j_i\,\tilde\varphi_j-\sigma_i.
\tag4.19
$$
Due to the relationships \thetag{4.19}, we can use the quantities
$\varphi_i$ in order to construct the components of affine
connection on the base of quantities \thetag{2.5}:
$$
\varGamma^k_{ij}=\theta^k_{ij}-\frac{\varphi_i\,\delta^k_j+
\varphi_j\,\delta^k_i}{3}
\tag4.20
$$\par
     In the case of general position $F\neq 0$. Therefore
we can use this field to construct the pair of vector fields
$\bold X$ and $\bold Y$ with the components:
$$
\xalignat 2
&X^i=\frac{\alpha^i}{F^2},
&&Y^i=\frac{\beta^i}{F^4}.
\tag4.21
\endxalignat
$$
Vectorial fields \thetag{4.21} form the moving frame on
the plane. Let's calculate the components of connection
\thetag{4.20} related to this frame. They are defined
as the coefficients in the following expansions:
$$
\xalignat 2
&\nabla_{\bold X}\bold X=\Gamma^1_{11}\,\bold X+
\Gamma^2_{11}\,\bold Y,
&
&\nabla_{\bold X}\bold Y=\Gamma^1_{12}\,\bold X+
\Gamma^2_{12}\,\bold Y,\\
\vspace{-1.7ex}
&&&\tag4.22\\
\vspace{-1.7ex}
&\nabla_{\bold Y}\bold X=\Gamma^1_{21}\,\bold X+
\Gamma^2_{21}\,\bold Y,
&
&\nabla_{\bold Y}\bold Y=\Gamma^1_{22}\,\bold X+
\Gamma^2_{22}\,\bold Y.
\endxalignat
$$
In contrast to the quantities $\varGamma^k_{ij}$ in \thetag{4.20},
the coefficients $\Gamma^k_{ij}$ from \thetag{4.22} do not
change by the point transformations \thetag{4.12}. These are
scalar fields, {\it i\. e\. scalar invariants} for the differential
equation \thetag{1.1}. Let's numerate them as follows: \linebreak
$I_1=\Gamma^1_{11}$, $I_2=\Gamma^2_{11}$, $I_3=\Gamma^1_{12}$,
\dots, $I_8=\Gamma^2_{22}$. The list of scalar invariants of
the equation \thetag{1.1} can be continued. By differentiating
these invariants along the vector fields $\bold X$ and $\bold Y$,
we obtain some new invariants $I_9=\bold XI_1$, \dots,
$I_{16}=\bold XI_8$, $I_{17}=\bold YI_1$, \dots, $I_{24}=\bold YI_8$.
Repeating this procedure, we shall get $16$ new invariants in each
step. The global structure of all these invariants $I_1$, $I_2$, $I_3$,
\dots determines the branching into three different cases:
\roster
\item in the infinite series of invariants $I_k(x,y)$ one can
      find two functionally independent ones;
\item invariants $I_k(x,y)$ are functionally dependent, but not
      all of them are constants;
\item all invariants $I_k(x,y)$ are constants.
\endroster
\par
      Let's write down the explicit expressions for the first
eight invariants in the special coordinates, where the conditions
\thetag{4.4} hold:
$$
\xalignat 2
&I_2=\frac{1}{3},
&&I_6=\frac{1}{3}\,\frac{F_{\sssize 1.0}}{F^3}.
\tag4.23
\endxalignat
$$
Formulas for $I_3$ and $I_8$ are a little more huge:
$$
\aligned
&I_3=3\,\frac{Q_{\sssize 1.0}}{F^4}-
14\,\frac{Q\,F_{\sssize 1.0}}{F^5}
+3\,\frac{Q^2}{F^4}+3\,F\,R+F_{\sssize 0.1},\\
\vspace{1ex}
&I_8=5\,\frac{Q\,F_{\sssize 1.0}}{F^5}-3\,\frac{Q^2}{F^4}
+5\,F_{\sssize 0.1}-3\,F\,R.
\endaligned
\tag4.24
$$
Formula for the invariant $I_7$ is even more huge than previous
ones:
$$
\aligned
I_7=9\,&\frac{Q\,Q_{1, 0}}{F^6}-45\,\frac{Q^2\,F_{1, 0}}{F^7}
+18\,\frac{Q^3}{F^6}+\\
\vspace{1ex}
&+27\,\frac{Q\,R}{F}-45\,\frac{Q\,F_{0, 1}}{F^2}
+9\,\frac{Q_{0, 1}}{F}+9\,F^4\,S.
\endaligned
\tag4.25
$$
For calculating other invariants $I_1$, $I_4$, and $I_5$ we
needn't special formulas different from \thetag{4.23},
\thetag{4.24}, and \thetag{4.25}. They are expressed through
$I_6$ and $I_8$:
$$
\xalignat 3
&\quad I_1=-4\,I_6,&&I_4=4\,I_6,&&I_5=-I_8.
\tag4.26
\endxalignat
$$
Due to the relationships \thetag{4.26}, the number of basic
invariants in the series $I_1$, $I_2$, $I_3$, $I_4$, \dots
is equal to four. These are $I_3$, $I_6$, $I_7$ and $I_8$.
Other invariants can be derived from the basic ones.
\head
5. Point symmetries in the case of general position.
\endhead
     In the case of general position $P=-F^5\neq 0$. Therefore
the equations \thetag{4.7} and \thetag{4.10} can be solved with
respect to the derivatives $U_{\sssize 01}$ and $V_{\sssize 0.1}$.
This gives
$$
\aligned
&U_{\sssize 0.1}=\frac{F_{\sssize 1.0}}{F}\,V+
\frac{F_{\sssize 0.1}}{F}\,U,\\
\vspace{1ex}
&V_{\sssize 0.1}=
\frac{Q_{\sssize 1.0}\,F-2\,F_{\sssize 1.0}\,Q}{F^6}\,V+
\frac{Q_{\sssize 0.1}\,F-2\,F_{\sssize 0.1}\,Q}{F^6}\,U.
\endaligned
\tag5.1
$$
The equations \thetag{5.1} are complemented by the equations
\thetag{4.5}, which are written in the the following form now:
$$
\xalignat 2
&U_{\sssize 1.0}=0,&&V_{\sssize 1.0}=
-\frac{2\,F_{\sssize 1.0}}{F}\,V-
\frac{2\,F_{\sssize 0.1}}{F}\,U.
\tag5.2
\endxalignat
$$
From \thetag{5.1} and \thetag{5.2} we derive a series of new
compatibility conditions:
$$
\xalignat 2
&U_{\sssize 1.1}=\frac{\partial U_{\sssize 0.1}}{\partial x},
&&U_{\sssize 0.2}=\frac{\partial U_{\sssize 0.1}}{\partial y},
\tag5.3\\
\vspace{1ex}
&V_{\sssize 1.1}=\frac{\partial V_{\sssize 0.1}}{\partial x},
&&V_{\sssize 0.2}=\frac{\partial V_{\sssize 0.1}}{\partial y}.
\tag5.4
\endxalignat
$$
The four relationships \thetag{5.3} and \thetag{5.4} can be
written in terms of the scalar invariants $I_3$, $I_6$, $I_7$
and $I_8$ considered above:
$$
\xalignat 2
&\frac{\partial I_3}{\partial x}\,V+\frac{\partial I_3}
{\partial y}\,U=0,
&&\frac{\partial I_6}{\partial x}\,V+\frac{\partial I_6}
{\partial y}\,U=0,\\
\vspace{-1ex}
&&&\tag5.5\\
\vspace{-1ex}
&\frac{\partial I_7}{\partial x}\,V+\frac{\partial I_7}
{\partial y}\,U=0,
&&\frac{\partial I_8}{\partial x}\,V+\frac{\partial I_8}
{\partial y}\,U=0.
\endxalignat
$$
Next four second order compatibility conditions, which are
not written in \thetag{5.3} and \thetag{5.4}, add nothing
new to the equations \thetag{5.5}.\par
     One more immediate consequence of the equations \thetag{5.1}
and \thetag{5.2} is the permutability of the vector fields
$\bold X$ and $\bold Y$ (defined by \thetag{4.21}) with the
vector field $\bold Z$ of the point symmetry for the equation
\thetag{1.1}:
$$
\xalignat 2
&[\bold Z,\bold X]=0,&&[\bold Z,\bold Y]=0.
\tag5.6
\endxalignat
$$\par
    The equations \thetag{5.5} have transparent geometrical
interpretation. They mean that basic invariants $I_3$, $I_6$,
$I_7$, and $I_8$ are constants along the integral lines
of the vector field of point symmetry $\bold Z$. From \thetag{4.26}
it follows that the same assertion is true for $I_1$, $I_4$, and
$I_5$. For the invariant $I_2$ this is trivial, since $I_2$ is
a constant. As a result, we conclude:
$$
\bold ZI_k=0\text{ \ for \ }k=1,\,\ldots,\,8.
\tag5.7
$$
Using the permutability equations \thetag{5.6}, we can expand
the relationships \thetag{5.7} for the whole infinite series
of invariants $I_k$. Let's do this by induction. Remember
that $I_{k+8}=\bold XI_k$ and $I_{k+16}=\bold YI_k$. Then
$$
\aligned
 &\bold ZI_{k+8}=\bold Z\bold XI_k=[\bold Z,\bold X]I_k+
\bold X\bold ZI_k=0,\\
\vspace{1ex}
&\bold ZI_{k+16}=\bold Z\bold YI_k=[\bold Z,\bold Y]I_k+
\bold Y\bold ZI_k=0.
\endaligned
$$
As an immediate consequence of the relationships $\bold ZI_k=0$,
for the whole series of invariants we get the following theorem.
\proclaim{Theorem 5.1} In the case of general position $F\neq 0$
the algebra of point symmetries for the equation \thetag{1.1}
with functionally independent invariants is trivial.
\endproclaim
     Indeed, the relationships $\bold ZI_p=0$ and $\bold ZI_q=0$
for two functionally independent invariants $I_p$ and $I_q$
lead to the vanishing of the vector field $\bold Z=0$.\par
\proclaim{Theorem 5.2} In the case of general position $F\neq 0$,
when all invariants $I_k$ are functionally dependent, but not all
of them are constants, the algebra of point symmetries of the
equation \thetag{1.1} is unidimensional.
\endproclaim
\demo{Proof} In the case of functionally dependent invariants
all equations $\bold ZI_k=0$ are the consequences of one of
them. Let's consider only the equations \thetag{5.1} and
\thetag{5.2}, forgetting for a while all other equations.
They form complete system of linear Pfaff equations.
Let's impose an extra restriction $\bold ZI_k=0$ on
them, where $I_k$ is some nonconstant invariant. Thus
we obtain the complete system of linear Pfaff equations
with restriction. Theory of such equations is given in
the Appendix B (see section 14 below). According to this
theory, compatibility condition for the equations
\thetag{5.1} and \thetag{5.2} is formed by the following
equations
$$
\xalignat 2
&\bold ZI_6=0,&&\bold ZI_3+\bold ZI_8=0,
\tag5.8
\endxalignat
$$
which hold due to the restriction $\bold ZI_k=0$. Therefore
the equations \thetag{5.1} and \thetag{5.2} with restriction
$\bold ZI_k=0$ has unidimensional space of solutions.\par
     Second order equations \thetag{5.4}, \thetag{5.3},
\thetag{4.11}, \thetag{4.8}, and \thetag{2.14} are compatible
with \thetag{5.1} and \thetag{5.2}. Therefore the dimension of
the algebra of point symmetries of the equation \thetag{1.1}
in this case is equal to one.\qed\enddemo
    In case of identically constant invariants the equations
\thetag{5.5} are \thetag{5.8} identically zero and we have
no restrictions for the Pfaff equations \thetag{5.1} and
\thetag{5.2}. They are compatible and the space of their
solutions is two-dimensional.
\proclaim{Theorem 5.3} In the case of general position $F\neq 0$,
when all invariants $I_k$ are constant, the algebra of point
symmetries of the equation \thetag{1.1} is two-dimensional.
\endproclaim
     For the arbitrary values of invariants this two-dimensional
Lie algebra is integrable but not commutative. However, if
$$
\xalignat 2
&I_6=0,&&I_3+I_8=0,
\tag5.9
\endxalignat
$$
then this algebra is Abelian. The relationships \thetag{5.9} are
derived by direct computation of structural constants.
\head
6. First case of intermediate degeneration.
\endhead
    In previous section we chose special coordinates, where
the conditions \thetag{4.4} are fulfilled, and we considered
the case of general position, when $P\neq 0$. Now let's
suppose that $P=0$. In this case, equation \thetag{4.7}
is identically zero, and the equation \thetag{4.10} is rewritten
as follows:
$$
2\,Q\,U_{\sssize 0.1}=Q_{\sssize 1.0}\,V+Q_{\sssize 0.1}\,U.
\tag6.1
$$
The solvability of the equation \thetag{6.1} with respect
to the derivative $U_{\sssize 0.1}$ is determined by the
condition $Q\neq 0$. For the case $P=0$, the quantities
\thetag{4.19} and fields \thetag{4.21} are not defined.
This require another theory of invariants  especially for
this case.\par
     Lets consider the point transformations \thetag{4.12}.
From the relationship \thetag{4.14} we see that the condition
$P=0$ is preserved by these transformations. For $P=\tilde P=0$,
the quantity $Q$ is transformed as follows:
$$
Q=\frac{\tilde Q}{{h'}^2}=(\det T)^2\,\tilde Q.
\tag6.2
$$
Due to \thetag{6.2}, the condition $Q\neq 0$ is also preserved
by the transformations \thetag{4.12}. In this case the quantity
$Q$ behave as pseudoscalar field of the weight $2$.\par
     Let's solve the equation \thetag{6.1} with respect to the
derivative $U_{\sssize 0.1}$ and then consider the differential
consequences of it:
$$
\xalignat 2
&U_{\sssize 1.1}=\frac{\partial U_{\sssize 0.1}}{\partial x},
&&U_{\sssize 0.2}=\frac{\partial U_{\sssize 0.1}}{\partial y}.
\tag6.3
\endxalignat
$$
First of the relationships \thetag{6.3} does not contain the
derivative $V_{\sssize 0.1}$. It is written as:
$$
\frac{1}{2}\,\frac{Q_{\sssize 2.0}}{Q}\,V
-\frac{Q_{\sssize 1.0}^2}{Q^2}\,V
+\frac{1}{2}\,\frac{Q_{\sssize 1.1}}{Q}\,U
-\frac{Q_{\sssize 1.0}\,Q_{\sssize 0.1}}{Q^2}\,U=0.
\tag6.4
$$
The second relationship contains the derivative $V_{\sssize 0.1}$.
It looks like
$$
\aligned
&\left(\frac{Q_{\sssize 1.0}}{Q}-
\frac{12}{5}\,Q\right)\,V_{\sssize 0.1}
-\frac{1}{2}\,\frac{Q_{\sssize 0.1}\,Q_{\sssize 1.0}}{Q^2}\,V
+\frac{Q_{\sssize 1.1}}{Q}\,V
-\frac{6}{5}\,R_{\sssize 0.1}\,U-\\
&-\frac{1}{2}\,\frac{Q_{\sssize 0.1}^2}{Q^2}\,U
+\frac{Q_{\sssize 0.2}}{Q}\,U
-\frac{6}{5}\,R_{\sssize 1.0}\,V
-\frac{3}{5}\,\frac{R\,Q_{\sssize 1.0}}{Q}\,V
-\frac{3}{5}\,\frac{R\,Q_{\sssize 0.1}}{Q}\,U=0.
\endaligned
\tag6.5
$$
The solvability of the equation \thetag{6.5} with respect to
the derivative $V_{\sssize 0.1}$ depends on the following
quantity:
$$
M=Q_{\sssize 1.0}-\frac{12}{5}\,Q^2.
\tag6.6
$$
The condition $M\neq 0$ leads to $Q\neq 0$. This condition
$M\neq 0$, combined with $P=0$, determines {\it the first
case of intermediate degeneration}.\par
     By means of direct calculations one can check that
for $P=0$, the quantity $M$ in \thetag{6.6}, by point
transformations \thetag{4.12}, is transformed as the
pseudoscalar field of the weight $4$
:
$$
M=\frac{\tilde M}{{h'}^4}=(\det T)^4\,\tilde M.
\tag6.7
$$
Due to \thetag{6.7}, the condition $M\neq 0$ is preserved by
the transformations \thetag{4.12}. The quantity $M$ in this
section plays the same role as the quantity $P$ in the case
of general position. However, the weights of the fields $Q$
and $M$ are even. This prevent us from defining the field
of the unit weight by taking the root of the appropriate
power. This circumstance make more difficult the theory of
invariants in this case, but one invariant can be written
right now:
$$
I_1=\frac{M}{Q^2}.
\tag6.8
$$\par
     In order to construct an affine connection, in previous
section we used the quantities \thetag{4.18}. Here in place of
them we introduce the following ones:
$$
\xalignat 2
&\varphi_1=-\frac{6}{5}\,Q,&&\varphi_2=-\frac{3}{5}\,R.
\tag6.9
\endxalignat
$$
It is easy to check that the quantities \thetag{6.9} obey the
rule \thetag{4.19} under the point transformations \thetag{4.12}.
This lets us construct an affine connection by formula
$$
\varGamma^k_{ij}=\theta^k_{ij}-\frac{\varphi_i\,\delta^k_j+
\varphi_j\,\delta^k_i}{3},
\tag6.10
$$
where the quantities $\theta^k_{ij}$ are defined according to
\thetag{2.4}. Despite the coincidence of formulas \thetag{6.10}
and \thetag{4.20} they determine two different connections.
This difference is due to the difference in parameters $\varphi_i$.
\par
     The quantities \thetag{6.9} and the components of affine
connection let us define the operation of covariant
differentiation for the pseudotensorial fields. In case of the
field of type $(r,s)$ and weight $m$ we set
$$
\aligned
\nabla_kF^{i_1\ldots\,i_r}_{j_1\ldots\,j_s}&=
\frac{\partial F^{i_1\ldots\,i_r}_{j_1\ldots\,j_s}}{\partial x^k}+
\sum^r_{n=1}\sum^2_{v_n=1}\Gamma^{i_n}_{k\,v_n}\,
F^{i_1\ldots\,v_n\ldots\,i_r}_{j_1\ldots\,j_s}-\\
&-\sum^s_{n=1}\sum^2_{w_n=1}\Gamma^{w_n}_{k\,j_n}\,
F^{i_1\ldots\,i_r}_{j_1\ldots\,w_n\ldots\,j_s}+
m\,\varphi_k\,F^{i_1\ldots\,i_r}_{j_1\ldots\,j_s}.
\endaligned
\tag6.11
$$
For $m=0$, this formula \thetag{6.11} coincides with the standard
formula for covariant derivatives (more details see in \cite{27}).
We apply this operation of covariant differentiation to the
pseudovectorial field $\alpha$ of the weight $2$ with the components
\thetag{4.4} and to the pseudovectorial field $\gamma$ with the
following components
$$
\xalignat 2
&\gamma^1=-2\,R_{\sssize 1.0}+3\,Q_{\sssize 0.1}+
\frac{6}{5}\,Q\,R,
&&\gamma^2=M.
\tag6.12
\endxalignat
$$
The condition of non-collinearity of the fields $\alpha$ and
$\gamma$ is exactly the condition $M\neq 0$, which is fulfilled
for the present case. Therefore we can write the expansions
analogous to \thetag{4.22}:
$$
\xalignat 2
&\nabla_{\alpha}\alpha=\Gamma^1_{11}\,\alpha+
\Gamma^2_{11}\,\gamma,
&
&\nabla_{\alpha}\gamma=\Gamma^1_{12}\,\alpha+
\Gamma^2_{12}\,\gamma,\\
\vspace{-1.7ex}
&&&\tag6.13\\
\vspace{-1.7ex}
&\nabla_{\gamma}\alpha=\Gamma^1_{21}\,\alpha+
\Gamma^2_{21}\,\gamma,
&&\nabla_{\gamma}\gamma=\Gamma^1_{22}\,\alpha+
\Gamma^2_{22}\,\gamma.
\endxalignat
$$
The field $\gamma$ with components \thetag{6.12} is of weight
$3$. This determines the weights of the pseudoscalar fields
$\Gamma^k_{ij}$ in the expansions \thetag{6.13}:
the field $\Gamma^2_{11}$ has the weight $1$, fields
$\Gamma^1_{11}$, $\Gamma^2_{12}$, and $\Gamma^2_{21}$ are of
weight $2$, fields $\Gamma^1_{12}$, $\Gamma^1_{21}$, and
$\Gamma^2_{22}$ have the weight $3$, and finally, the field
$\Gamma^1_{22}$ has the weight $4$. Now we give explicit formulas
for these fields. Part of them are very simple
$$
\xalignat 2
&\Gamma^2_{11}=0,&&\Gamma^1_{11}=
\Gamma^2_{21}=-\frac{3}{5}\,Q,\\
\vspace{-1ex}
&&&\tag6.14\\
\vspace{-1ex}
&\Gamma^1_{21}=0,&&\Gamma^2_{12}=\frac{M_{\sssize 1.0}}{M}
-\frac{21}{5}\,Q.
\endxalignat
$$
Formulas for the pair of fields $\Gamma^1_{12}$ and $\Gamma^2_{22}$
differ from each other only by sign:
$$
\aligned
\Gamma^2_{22}=-\Gamma^1_{12}&=M_{\sssize 0.1}
-\frac{72}{5}\,Q\,Q_{\sssize 0.1}+\frac{48}{5}\,Q\,R_{\sssize 1.0}
-\frac{12}{5}\,R\,M-\\
\vspace{1ex}
&-\frac{144}{25}\,R\,Q^2
-2\,\frac{R_{\sssize 1.0}\,M_{\sssize 1.0}}{M}
+3\,\frac{Q_{\sssize 0.1}\,M_{\sssize 1.0}}{M}
+\frac{6}{5}\,\frac{Q\,R\,M_{\sssize 1.0}}{M}.
\endaligned
\tag6.15
$$
Formula for $\Gamma^1_{22}$ appears to be the most huge. It has
the form:
$$
\aligned
\Gamma^1_{22}&=\frac{24}{5}\,\frac{R_{\sssize 1.0}\,
Q\,R\,M_{\sssize 1.0}}{M}
-\frac{36}{5}\,\frac{Q_{\sssize 0.1}\,Q\,R\,M_{\sssize 1.0}}{M}
-2\,M\,R_{\sssize 1.1}+\\
\vspace{1ex}
&+3\,M\,Q_{\sssize 0.2}
+M^2\,S
-4\,\frac{R_{\sssize 1.0}^2\,M_{\sssize 1.0}}{M}
-9\,\frac{Q_{\sssize 0.1}^2\,M_{\sssize 1.0}}{M}-\\
\vspace{1ex}
&-\frac{324}{5}\,R_{\sssize 1.0}\,Q\,Q_{\sssize 0.1}
-\frac{648}{25}\,R_{\sssize 1.0}\,R\,Q^2
+\frac{972}{25}\,Q_{\sssize 0.1}\,R\,Q^2-\\
\vspace{1ex}
&-\frac{12}{5}\,Q\,R\,M_{\sssize 0.1}
-\frac{36}{25}\,\frac{Q^2\,R^2\,M_{\sssize 1.0}}{M}
+12\,\frac{R_{\sssize 1.0}\,Q_{\sssize 0.1}\,M_{\sssize 1.0}}{M}+\\
\vspace{1ex}
&+\frac{126}{25}\,Q\,R^2\,M
-\frac{42}{5}\,R_{\sssize 1.0}\,R\,M
+\frac{69}{5}\,Q_{\sssize 0.1}\,R\,M+\\
\vspace{1ex}
&+\frac {6}{5}\,M\,Q\,R_{\sssize 0.1}
+4\,R_{\sssize 1.0}\,M_{\sssize 0.1}
+\frac{108}{5}\,Q\,R_{\sssize 1.0}^2-\\
\vspace{1ex}
&-6\,Q_{\sssize 0.1}\,M_{\sssize 0.1}
+\frac{243}{5}\,Q\,Q_{\sssize 0.1}^2
+\frac{972}{125}\,Q^3\,R^2.
\endaligned
\tag6.16
$$\par
     The quantities $\varphi_i$ themselves determine the
skew-symmetric tensor field $\omega_{ij}$ of the weight zero
with the following components:
$$
\omega_{ij}=\frac{\partial\varphi_i}{\partial x^j}-
\frac{\partial\varphi_j}{\partial x^i}.
\tag6.17
$$
Upon contracting \thetag{6.17} with the matrix $d^{ij}$ from
\thetag{2.6}, we get the pseudoscalar field $\Omega$ of the
following form
$$
\Omega=\frac{5}{6}\sum^2_{i=1}\sum^2_{j=1}\omega_{ij}\,d^{ij}=
R_{\sssize 1.0}-2\,Q_{\sssize 0.1}.
\tag6.18
$$
Field \thetag{6.18} has the weight $1$. This field is used in
order to determine one more scalar invariant of the equation
\thetag{1.1}:
$$
I_2=\frac{\Omega^2}{Q}=\frac{R_{\sssize 1.0}^2-4\,R_{\sssize 1.0}
\,Q_{\sssize 0.1}+4\,Q_{\sssize 0.1}^2}{Q}.
\tag6.19
$$
Third invariant $I_3$ is determined by the field $\Gamma^1_{22}$
from \thetag{6.16}. It has the form:
$$
I_3=\frac{\Gamma^1_{22}\,Q^2}{M^2}.
\tag6.20
$$
Invariants \thetag{6.8}, \thetag{6.19}, and \thetag{6.20} are
basic ones. Other invariants are derived from them by means
of differentiation along the fields $\alpha$ and $\gamma$
according to the rule:
$$
\xalignat 2
&I_{k+3}=\frac{\nabla_\alpha I_k}{Q},
&&I_{k+6}=\frac{\bigl(\nabla_\gamma I_k\bigr)^2}{Q^3}.
\tag6.21
\endxalignat
$$
Applying the rule \thetag{6.21} repeatedly step by step, we
get 6 new invariants in each step. Moreover, we have:
$$
\gather
I_1\,\Gamma^2_{12}=I_4\,Q-\frac{3}{5}\,I_1\,Q-2\,I_1^2\,Q,
\tag6.22\\
\vspace{2ex}
\aligned
 \bigl(I_1\,\Gamma^2_{22}\bigr)^4&+\bigl(I_7\,Q^3\bigr)^2
 +\bigl(16\,I_2\,Q^3\,{I_1}^4\bigr)^2=\\
 \vspace{1ex}
 &=32\,I_7\,Q^6\,I_2\,{I_1}^4
 +2\,\bigl(I_7\,Q^3+16\,I_2\,Q^3\,{I_1}^4\bigr)\,
 \bigl(I_1\,\Gamma^2_{22}\bigr)^2.\hskip-2em
\endaligned
\tag6.23
\endgather
$$
The relationships \thetag{6.22} and \thetag{6.23} bind the
invariants $I_4$ and $I_7$ with the fields $\Gamma^1_{12}$
and $\Gamma^2_{22}$ from \thetag{6.14} and \thetag{6.15}.\par
     Now we return to the study of the point symmetries of the
equation \thetag{1.1} taking into account the above theory of
invariants. Using the notation \thetag{6.6} and keeping in mind
that $M\neq 0$, we solve the equation \thetag{6.5} with respect
to the derivative $V_{\sssize 0.1}$. As a result, we complement
the equation \thetag{6.4} with two differential consequences
of the equation \thetag{6.5}. They have the form \thetag{5.4}.
These two equations can be reduced to the following relationships
for the invariants $I_2$ and $I_3$:
$$
\xalignat 2
&\quad\frac{\partial I_2}{\partial x}\,V+\frac{\partial I_2}
{\partial y}\,U=0,
&&\frac{\partial I_3}{\partial x}\,V+\frac{\partial I_3}
{\partial y}\,U=0.
\tag6.24
\endxalignat
$$
The equation \thetag{6.4} itself can be transformed to the
analogous relationship for $I_1$:
$$
\quad\frac{\partial I_1}{\partial x}\,V+\frac{\partial I_1}
{\partial y}\,U=0.
\tag6.25
$$
From \thetag{6.24} and \thetag{6.25} one can derive similar
relationships for all invariants in the series $I_1$, $I_2$, $I_3$,
\dots constructed according to the rule \thetag{6.21}.
Proof of this fact is based on the relationship \thetag{5.6}
for the following two vector fields:
$$
\xalignat 2
&\bold X=\frac{\alpha}{Q},
&&\bold Y=\frac{\gamma}{Q^{3/2}}.
\endxalignat
$$
The relationships \thetag{5.6} for the above fields are
proved by direct calculations.\par
     As in case of general position, here we have three
different subcases due to the general structure of invariants
in the sequence $I_1$, $I_2$, $I_3$, \dots, they are
\roster
\item the case, when in the infinite series of invariants
      $I_k(x,y)$ one can find two functionally independent ones;
\item the case, when invariants $I_k(x,y)$ are functionally
      dependent, but not all of them are constants;
\item the case, when all invariants $I_k(x,y)$ are constants.
\endroster
\par
    In the first case algebra of point symmetries of the
equation \thetag{1.1} is trivial, in the second case this
algebra is unidimensional. And in the last case it is
two-dimensional. The commutativity conditions for this
algebra is written as
$$
\xalignat 2
&I_1=-\frac{12}{5},
&&I_2=0.
\tag6.26
\endxalignat
$$
When at least one of these conditions \thetag{6.26} is broken,
the algebra of symmetries is integrable, but not abelian.\par
\head
7. Second case of intermediate degeneration.
\endhead
     This case is determined by the conditions $P=-F^5=0$,
$Q\neq 0$, and $M=0$ in the special coordinates, where the
conditions \thetag{4.4} are fulfilled. From $M=0$ we have the
differential equation $Q_{\sssize 1.0}=12/5\,Q^2$ for the
function $Q(x,y)$. This differential equation is easily
solvable:
$$
Q=-\frac{5}{12\,x+c(y)}.
\tag7.1
$$
Upon doing the point transformation \thetag{4.12} with the proper
choice of the function $g(y)$ in it, the field $Q$ from \thetag{7.1}
can be brought to the form
$$
Q=-\frac{5}{12\,x}.
\tag7.2
$$
Then point transformations \thetag{4.12} that preserve the form
\thetag{7.2} of the field $Q$ are determined as follows:
$$
\xalignat 2
&\tilde x=\frac{x}{h'(y)^2},
&&\tilde y=h(y).
\tag7.3
\endxalignat
$$
By substituting \thetag{7.2} into the equation $A=0$ from
\thetag{4.4}, we derive the differential equation for the
function $R(x,y)$:
$$
R_{\sssize 2.0}=-\frac{5}{4}\,\frac{R_{\sssize 1.0}}{x}.
\tag7.4
$$
This equation \thetag{7.4} is also easily solvable. Its general
solution has the form
$$
R=r(y)+c(y)\,|x|^{-1/4}.
\tag7.5
$$
Now let's consider the case, when $c(y)\neq 0$ in \thetag{7.5}.
Let's do the point transformation \thetag{7.3} and take into
account that $R$ is a pseudoscalar field of the weight $-1$
respective to such transformations: $R=(\det T)^{-1}\,\tilde R$.
At the expense of proper choice of $h(y)$, the field \thetag{7.5}
can be brought to the following more simple form:
$$
R=r(y)+|x|^{-1/4}.
\tag7.6
$$
Point transformations \thetag{7.3} that preserve the form
\thetag{7.6} for $R$ are extremely simple:
$$
\xalignat 2
&\tilde x=x,
&&\tilde y=y+\const.
\tag7.7
\endxalignat
$$
All nonzero coefficients $Q$, $R$ and $S$ of the equation
\thetag{1.1} form the scalar fields (scalar invariants) with
respect to the transformations \thetag{7.7}.\par
     Let's substitute \thetag{7.2} and \thetag{7.6} into
the equation $B=1$ from \thetag{4.4}. As a result, we get
the differential equation for the function $S(x,y)$:
$$
S_{\sssize 2.0}
-\frac{5\,S_{\sssize 1.0}}{4\,x}
+\frac{5\,S}{4\,x^2}=1
-\frac{3\,r(y)}{2\,x\,|x|^{1/4}}
-\frac{3}{2\,x\,|x|^{1/2}}.
\tag7.8
$$
General solution for the linear ordinary differential equation
\thetag{7.8} can be written in the following explicit form:
$$
S=\sigma(y)\,|x|^{5/4}-4\,s(y)\,x
+\frac{4}{3}\,|x|^2-
12\,\frac{r(y)\,|x|^{7/4}}{x}-4\,\frac{|x|^{3/2}}{x}.
\tag7.9
$$
Let's substitute \thetag{7.2}, \thetag{7.4} and \thetag{7.9}
into the equations \thetag{6.4} and \thetag{6.5} for the
field of point symmetry. Thereby the equation \thetag{6.4}
appears to become an identity. The equation \thetag{6.5}
then is rewritten as follows:
$$
\left(-\frac{3}{4}\,\frac{1}{|x|^{1/4}\,x}
-\frac{1}{2}\,\frac{r(y)}{x}\right)\,V
+r'(y)\,U=0.
\tag7.10
$$
Let's differentiate the equation \thetag{7.10} with respect
to the variable $x$ taking into account the relationships
\thetag{4.5} and \thetag{6.1}. As a result, we get the
equation
$$
-\frac{9}{80}\,\frac{V}{|x|^{1/4}\,x^2}=0,
$$
which leads to the vanishing of one of the components of the
vector field of point symmetry: $V=0$. By substituting $V=0$
into the equation \thetag{7.10}, we get $r'(y)\,U=0$. If the
function $r(y)$ is not constant $r'(y)\neq 0$, then $U=0$
and the algebra of point symmetries of the equation
\thetag{1.1} is trivial.\par
     In the case, when $r'(y)=0$, from $V=0$ we derive that
$V_{\sssize 0.1}=0$. This lets us write the pair of new
compatibility conditions:
$$
\xalignat 2
&\frac{\partial V}{\partial x}=V_{\sssize 1.0},
&&\frac{\partial V_{\sssize 0.1}}{\partial x}=V_{\sssize 1.1}.
\tag7.11
\endxalignat
$$
First of the relationships \thetag{7.11} holds identically, from
the second one we derive:
$$
\left(\sigma'(y)\,|x|^{5/4}-4\,s'(y)\,x\right)\,U=0.
\tag7.12
$$
As an immediate consequence of \thetag{7.12} we have the following
theorem.\pagebreak
\proclaim{Theorem 7.1} In the second case of intermediate
degeneration the algebra of point symmetries of the equation
\thetag{1.1} is unidimensional if and only if the parameters
$r(y)$, $s(y)$ and $\sigma(y)$ in \thetag{7.4} and \thetag{7.9}
are identically constant:
$$
\xalignat 3
&r'(y)=0,&&s'(y)=0,&&\sigma'(y)=0.
\tag7.13
\endxalignat
$$
If at least one of the conditions \thetag{7.13} is broken, then
the algebra of point symmetries is trivial.
\endproclaim
\head
8. Third case of intermediate degeneration.
\endhead
     This case splits off from the second case of intermediate
degeneration by the conditions $c(y)=0$ and $r(y)\neq 0$ in the
formula \thetag{7.4}. By the proper choice of the function
$h(y)$ in the point transformation \thetag{7.3}, upon applying
such transformation, we can get the following relationship:
$$
R=1.
\tag8.1
$$
Due to \thetag{8.1}, the equation \thetag{7.8} is replaced by
the following more simple one:
$$
S_{\sssize 2.0}
-\frac{5\,S_{\sssize 1.0}}{4\,x}
+\frac{5\,S}{4\,x^2}=1.
\tag8.2
$$
The general solution for the differential equation \thetag{8.2}
can be written explicitly:
$$
S=\sigma(y)\,|x|^{5/4}-4\,s(y)\,x
+\frac{4}{3}\,|x|^2.
\tag8.3
$$
Let's substitute \thetag{7.2}, \thetag{8.1} and \thetag{8.3}
into the equations \thetag{6.4} and \thetag{6.5} for the
field of point symmetry. Then the equation \thetag{6.4}
appears to become an identity. The equation \thetag{6.5}
reduces to the following relationship
$$
-\frac{1}{2}\,\frac{V}{x}=0,
$$
which leads to the vanishing of one of the components of the
vector field of point symmetry: $V=0$. Now we can rewrite
the compatibility conditions \thetag{7.11}, one of them is
an identity, another coincides with \thetag{7.12}.
\proclaim{Theorem 8.1} In the third case of intermediate
degeneration the algebra of point symmetries of the equation
\thetag{1.1} is unidimensional if and only if the parameters
$s(y)$ and $\sigma(y)$ in \thetag{8.3} are identically
constant:
$$
\xalignat 2
&s'(y)=0,&&\sigma'(y)=0.
\tag8.4
\endxalignat
$$
If at least one of the conditions \thetag{8.4} is broken, then
the algebra of point symmetries is trivial.
\endproclaim
\head
9. Fourth case of intermediate degeneration.
\endhead
     This case splits off from the second case of intermediate
degeneration by the condition of simultaneous vanishing $c(y)=0$
and $r(y)=0$ in the formula \thetag{7.4}. This is equivalent to
the vanishing of $R$:
$$
R=0.
\tag9.1
$$
The equation \thetag{8.2} for the function $S(x,y)$ in this
case remains unchanged. Therefore $S$ is defined by the
formula \thetag{8.3}. However, in contrast to \thetag{8.1},
the condition \thetag{9.1} specify no subclass in the class
of point transformations \thetag{7.3}. Therefore we can use
such transformations for to simplify the formula \thetag{8.3}.
Transformation rule for $S$ by the point transformations
\thetag{7.3} has the form
$$
S=3\,\frac{x\,{h''}^2}{{h'}^2}
-2\,\frac{x\,h'''}{h'}
+{h'}^4\,\tilde S.
\tag9.2
$$
From \thetag{9.2} we derive the following transformation rules
for $\sigma(y)$ and $s(y)$:
$$
\aligned
&\sigma(y)=\tilde\sigma(\tilde y)\,|h'(y)|^{3/2},\\
\vspace{1ex}
&s(y)=\tilde s(\tilde y)\,h'(y)^2-\frac{3}{4}\,
\frac{{h''(y)}^2}{{h'(y)}^2}
+\frac{1}{2}\,\frac{h'''(y)}{h'(y)}.
\endaligned
\tag9.3
$$
From \thetag{9.3} we can see that by means of the proper
choice of $h(y)$ in \thetag{7.3} we can transform $S(x,y)$
to the form
$$
S=\sigma(y)\,|x|^{5/4}+\frac{4}{3}\,|x|^2.
\tag9.4
$$
Now we are able to specify the subclass of point transformations
\thetag{7.3} that preserve the form \thetag{9.4} of the field
$S$. Function $h(y)$ for such transformations should satisfy
the differential equation $3\,{h''}^2=2\,h'''\,h'$. This is
linear-fractional function
$$
h(y)=\frac{a\,y+b}{c\,y+d}.
$$\par
     In order to calculate the algebra of point symmetries
for the equation \thetag{1.1} in the present case, let's
substitute \thetag{7.2}, \thetag{9.1}, and \thetag{9.4}
into the equations \thetag{6.4} and \thetag{6.5}. Both of
them appear to be satisfied identically. This means that
for to obtain nontrivial equations for the components of
the vector field of point symmetry, in this case, we should
return to the equations \thetag{2.14} and to the
relationships \thetag{2.15}--\thetag{2.22}. From the
following two compatibility conditions
$$
\xalignat 2
&\frac{\partial U_{\sssize 0.2}}{\partial y}=
U_{\sssize 0.3},
&&\frac{\partial V_{\sssize 1.1}} {\partial y}=
V_{\sssize 1.2},
\endxalignat
$$
we obtain one equation binding $V$ and $U$. It looks like
$$
-3\,\sigma(y)\,V+4\,x\,\sigma'(y)\,U=0.
\tag9.5
$$
Other compatibility conditions in the third order of
derivatives appears to be fulfilled identically.\par
     Let's add the condition $\sigma(y)\neq 0$ to \thetag{9.1}.
These two conditions, together with $P=0$, $Q\neq 0$, and
\thetag{4.4}, are the very conditions that specify the fourth
case of intermediate degeneration. Using $\sigma(y)\neq 0$,
we solve the equation \thetag{9.5} with respect to $V$
and then calculate the derivative
$$
V_{\sssize 0.1}=\frac {4}{3}\,\frac{x\,\sigma''}{\sigma}\,U
-\frac{20}{9}\,\frac{x\,{\sigma'}^2}{\sigma^2}\,U.
\tag9.6
$$
Now we can consider the following differential consequence
from \thetag{9.6}:
$$
\frac{\partial V_{\sssize 0.1}}{\partial y}=
U_{\sssize 0.2}.
\tag9.7
$$
The relationship \thetag{9.7} is written as the equation for
$U$:
$$
\left(\sigma'''-5\,\frac{\sigma''\,\sigma'}{\sigma}
+\frac{40}{9}\,\frac{{\sigma'}^3}{\sigma^2}\right)
\,U=0.
\tag9.8
$$
Other differential consequences add nothing to \thetag{9.8}
and \thetag{9.5}.
\proclaim{Theorem 9.1} In the fourth case of intermediate
degeneration the algebra of point symmetries of the equation
\thetag{1.1} is unidimensional if and only if the function
\linebreak $\sigma(y)\neq 0$ in \thetag{9.4} satisfies the
differential equation of the form:
$$
\sigma'''-5\,\frac{\sigma''\,\sigma'}{\sigma}
+\frac{40}{9}\,\frac{{\sigma'}^3}{\sigma^2}=0.
\tag9.9
$$
If the differential equation \thetag{9.9} does not hold, then
the algebra of point symmetries is trivial.
\endproclaim
\head
10. Fifth case of intermediate degeneration.
\endhead
     Fifth case splits off from the fourth case of intermediate
degeneration by the condition of vanishing $\sigma(y)=0$ in the
formula \thetag{9.4}. Then the equation \thetag{9.5} becomes
an identity and no more differential consequences are available.
But the differential consequences, which are already available
$$
\xalignat 2
&V_{\sssize 1.0}=\frac{1}{x}\,V,
&&V_{\sssize 0.2}=0,\\
\vspace{-1.2ex}
&&&
\tag10.1\\
\vspace{-1.2ex}
&U_{\sssize 1.0}=0,
&&U_{\sssize 0.1}=-\frac{1}{2\,x}\,V,
\endxalignat
$$
can be written as the system of Pfaff equations \thetag{2.23}.
In order to do this, we should construct the vector-column $\Psi$
of the following three functions: $V$, $U$, $V_{\sssize 0.1}$.
From $\sigma(y)=0$ we derive the compatibility of this complete
Pfaff system, i\. e\. the condition $\sigma(y)=0$ converts to
identity the matrix equations \thetag{2.24}. Compatible system
of Pfaff equations \thetag{10.1} is easily solvable:
$$
\xalignat 2
&V=-4\,a\,x\,y-4\,b\,x,
&&U=a\,y^2+2\,b\,y+c.
\tag10.2
\endxalignat
$$
Its general solution is parameterized by three arbitrary
constants $a$, $b$, and $c$. From \thetag{10.2} it is easy to
get the following characterization for base elements of the
algebra of point symmetries of the equation \thetag{1.1} for
the present case.\par
    1. {\it Shift of function value}: $x\mapsto  x$, $y\mapsto y+t$.
Components of corresponding vector field are: $V=0$, $U=1$.\par
    2. {\it Simultaneous blowing up of argument and the function
value}: $x\mapsto x\,e^{-2t}$, $y\mapsto y\,e^t$. Components of
corresponding vector field are: $V=-2\,x$, $U=y$.\par
    3. {\it Inversion of the function value with simultaneous
blowing up the argument}. Components of vector field are: $V=-2\,x\,y$,
$U=y^2/2$. The transformations, forming the one-parametric group,
have the form:
$$
\xalignat 2
&x\mapsto x\,\left(1-\frac{y\,t}{2}\right)^4,
&&y\mapsto\frac{2\,y}{2-y\,t}.
\endxalignat
$$\par
\proclaim{Theorem 10.1} The algebra of point symmetries of
the equation \thetag{1.1} in the fifth case of intermediate
degeneration is isomorphic to the matrix algebra
$\sll(2,\Bbb R)$.
\endproclaim
\demo{Proof} Arbitrary vector field from the algebra of
symmetries is defined by three constants $a$, $b$, and $c$
according to the formulas \thetag{10.2}. Let's consider
the map
$$
(a,b,c)\mapsto
\Vmatrix b & c\\
\vspace{1ex}
a &-b\endVmatrix.
\tag10.3
$$
By means of direct calculations one can prove that this
map \thetag{10.3} is the very map that establish the
isomorphism of the algebra of symmetries of the equation
\thetag{1.1} and the matrix algebra $\sll(2,\Bbb R)$.
\qed\enddemo
\head
11. Sixth case of intermediate degeneration.
\endhead
    In all five previous cases of intermediate degeneration
the field $Q$ was nonzero (provided the conditions \thetag{4.4}
and the equality $P=0$ are fulfilled). Now let's set $Q=0$.
Then the condition $A=0$ from \thetag{4.4} gives
$R_{\sssize 2.0}=0$. Therefore
$$
R=c(y)\,x+r(y).
\tag11.1
$$
{\it Sixth case of intermediate degeneration} is specified
by the additional condition $c(y)\neq 0$ in \thetag{11.1}.
The transformation rule for the quantity $R$ under the point
transformations \thetag{4.12} has the form:
$$
R=-\frac{5}{3}\,\frac{h''(y)}{h'(y)}+h'(y)\,\tilde R.
\tag11.2
$$
Due to \thetag{11.2}, at the expense of proper choice of
functions $h(y)$ and $g(y)$ in \thetag{4.12} we can
ensure the relationships $c(y)=1$ and $r(y)=0$ upon
applying the transformation \thetag{4.12}. Then for $R$
we have
$$
R=x.
\tag11.3
$$
Point transformations \thetag{4.12} that preserve the form of
the function $R$ in \thetag{11.3} form special subclass defined
by the formulas
$$
\xalignat 2
&\tilde x=x,
&&\tilde y=y+\const.
\tag11.4
\endxalignat
$$
Both nonzero coefficients $R$ and $S$ in this case are the
scalar fields (scalar invariants) respective to the
transformations \thetag{11.4}.\par
     Now let's consider the condition $B=1$ from \thetag{4.4}.
This condition leads to the differential equation for $S$:
$$
S_{\sssize 2.0}=6\,x+1.
\tag11.5
$$
One can easily write the general solution for \thetag{11.5}. It
has the following form:
$$
S=x^3+\frac{1}{2}\,x^2+\sigma(y)\,x+s(y).
\tag11.6
$$\par
     In order to calculate the algebra of point symmetries
for the equation \thetag{1.1}, let's substitute $P=0$ and
$Q=0$ into the equations \thetag{4.7} and \thetag{4.10}.
This trivialize both of them and we cannot express the
derivatives $U_{\sssize 0.1}$ and $V_{\sssize 0.1}$ from
them. Therefore we are to consider the differential consequences
of the third order derived from \thetag{2.14}, \thetag{4.8},
and \thetag{4.11}. The following four equations
$$
\xalignat 2
&\frac{\partial U_{\sssize 0.2}}{\partial x}=
U_{\sssize 1.2},
&&\frac{\partial U_{\sssize 1.1}}{\partial y}=
U_{\sssize 1.2},\\
\vspace{1ex}
&\frac{\partial V_{\sssize 1.1}}{\partial x}=
V_{\sssize 2.1},
&&\frac{\partial V_{\sssize 2.0}}{\partial y}=
V_{\sssize 2.1}
\endxalignat
$$
are brought to one equation of the very simple form:
$$
U_{\sssize 0.1}=0.
\tag11.7
$$
Taking into account \thetag{11.7}, from next pair of equations
of the third order
$$
\xalignat 2
&\frac{\partial U_{\sssize 0.2}}{\partial y}=
U_{\sssize 0.3},
&&\frac{\partial V_{\sssize 1.1}}{\partial y}=
V_{\sssize 1.2},
\endxalignat
$$
we derive one more equation that determine the derivative
$V_{\sssize 0.1}$:
$$
V_{\sssize 0.1}=\left(-\frac{14}{15}\,x-\frac{5}{9}\right)\,
V-\frac{5}{9}\,\sigma'(y)\,U.
\tag11.8
$$
Other compatibility equations of the third order add nothing
new. But nevertheless, using \thetag{11.7} and \thetag{11.8},
\pagebreak
we can come back to the second order compatibility conditions.
Now they are reduced to the pair of relationships, one of which
is $V=0$. Another one has the form:
$$
\left(s'(y)-\frac{4}{27}\,\sigma'(y)\,x+\frac{25}{81}\,
\sigma'(y)-\frac {5}{9}\,\sigma''(y)\right)\,U=0.
\tag11.9
$$
From $V=0$ and from \thetag{11.9} we can derive the following
theorem.
\proclaim{Theorem 11.1} In the sixth case of intermediate
degeneration the algebra of point symmetries of the equation
\thetag{1.1} is unidimensional if and only if the parameters
$s(y)$ and $\sigma(y)$ in \thetag{11.6} are identically
constant:
$$
\xalignat 2
&s'(y)=0,&&\sigma'(y)=0.
\tag11.10
\endxalignat
$$
If at least one of the conditions \thetag{11.10} is broken, then
the algebra of point symmetries is trivial.
\endproclaim
\head
12. Seventh case of intermediate degeneration.
\endhead
     This case splits off from sixth case of intermediate
degeneration by the condition $c(y)=0$ in the formula
\thetag{11.1}. Using the transformation rule \thetag{11.2},
in this case, we can provide the vanishing of the field $R$:
$$
R=0.
\tag12.1
$$
Point transformations of the form \thetag{4.12} that preserve
the condition \thetag{12.1} are defined by the following
formulas:
$$
\xalignat 2
&\tilde x=\frac{x}{a^2}+g(y),
&&\tilde y=a\,y+b.
\tag12.2
\endxalignat
$$
From the condition $B=1$ in \thetag{4.4}, we get the
differential equation $S_{\sssize 2.0}=1$ for the field $S$.
General solution of this trivial equation has the form:
$$
S=\frac{1}{2}\,x^2+\sigma(y)\,x+s(y).
\tag12.3
$$
Point transformations \thetag{12.2} transform the field $S$
according to the following rule
$$
S=a^2\,g''(y)+a^4\,\tilde S.
\tag12.4
$$
From \thetag{12.4} and from \thetag{12.2} one easily derive the
transformation rules for the coefficients $\sigma(y)$ and
$s(y)$ in \thetag{12.3}. They have the form:
$$
\aligned
&\sigma(y)=a^2\,\tilde\sigma(\tilde y)+a^2\,g(y),\\
&s(y)=a^4\,\tilde s(\tilde y)+a^2\,g''(y)+
a^4\,\tilde\sigma(\tilde y)\,g(y)+\frac{1}{2}\,a^4\,g(y)^2.
\endaligned
\tag12.5
$$
Now, from \thetag{12.5}, we see that, using the arbitrariness
in the choice of local coordinates, defined by the point
transformations \thetag{12.2}, we always can vanish $\sigma(y)$
in \thetag{12.3}. For $S$ from $\sigma(y)=0$ we get
$$
S=\frac{1}{2}\,x^2+s(y).
\tag12.6
$$
Point transformations \thetag{12.2} that preserve the form
of field $S$ in \thetag{12.6} form the subclass that doesn't
contain the functional parameters:
$$
\xalignat 2
&\tilde x=\frac{x}{a^2},
&&\tilde y=a\,y+b.
\tag12.7
\endxalignat
$$
The field $S$ is a pseudoscalar field of the weight $-4$
respective to the transformations \thetag{12.7}. But we
can easily construct the scalar field $I$:
$$
I=\frac{S}{x^2}.
$$\par
     Here, as in the previous case, the equations
\thetag{4.7} and \thetag{4.10} are identically zero
due to $P=0$ and $Q=0$. In order to find the algebra
of point symmetries, we should consider the differential
consequences of higher order. The following two
compatibility equations of the third order
$$
\xalignat 2
&\frac{\partial U_{\sssize 0.2}}{\partial y}=
U_{\sssize 0.3},
&&\frac{\partial V_{\sssize 1.1}}{\partial y}=
U_{\sssize 1.2}
\endxalignat
$$
are reduced to one relationship, defining the derivative
$U_{\sssize 0.1}$:
$$
U_{\sssize 0.1}=-\frac{1}{2\,x}\,V.
\tag12.8
$$
This relationship \thetag{12.8} gives rise to the new
compatibility conditions of the second order. Two of them
$$
\xalignat 2
&\frac{\partial U_{\sssize 0.1}}{\partial y}=
U_{\sssize 0.2},
&&\frac{\partial V_{\sssize 1.0}}{\partial y}=
U_{\sssize 1.1}
\endxalignat
$$
are brought to the relationship that leads to the vanishing
of the derivative $V_{\sssize 0.1}$:
$$
V_{\sssize 0.1}=0.
\tag12.9
$$
From \thetag{12.9} we derive the equation $V_{\sssize 0.2}=0$.
In explicit form this equation is written as:
$$
\frac{2}{x}\,s(y)\,V-s'(y)\,U=0.
\tag12.10
$$
When $s(y)=0$, the equation \thetag{12.10} is fulfilled
identically and no more equations are available. Therefore
we have the following theorem.
\proclaim{Theorem 12.1} In the seventh case of intermediate
degeneration the algebra of point transformations of the
equation \thetag{1.1} is two-dimensional if and only if
the function $s(y)$ in \thetag{12.6} is identically zero.
This algebra is integrable but it is not Abelian.
\endproclaim
    Now let's suppose that $s(y)\neq 0$. In this case
the equation \thetag{12.10} can be solved with respect
to $V$ and we can derive two differential consequences
of the first order from it. One of them has the form:
$$
\left(\frac{4\,s''(y)}{s(y)}-\frac{5\,s'(y)^2}{s(y)^2}
\right)\,U=0.
$$
\proclaim{Theorem 12.2} In the seventh case of intermediate
degeneration the algebra of point transformations of the
equation \thetag{1.1} is unidimensional if and only if
the function $s(y)$ in \thetag{12.6} is the solution of the
following differential equation:
$$
4\,s''(y)-\frac{5\,s'(y)^2}{s(y)}=0.
\tag12.11
$$
If the differential equation \thetag{12.11} does not hold, then
the algebra of point symmetries is trivial.
\endproclaim
\head
13. Appendix A. Systems of Pfaff equations.
\endhead
    Let's consider the system of $n$ functions: $u^1(x,y),\ldots,
u^n(x,y)$. The following two systems of equations form
{\it the complete system of Pfaff equations}:
$$
\xalignat 2
&\cases
 u^1_{\sssize 1.0}=U^1_{\sssize 1.0}(u^1,\ldots,u^n,x,y),\\
 \vspace{0.1ex}
 \hdotsfor 1\\
 \vspace{0.5ex}
 u^n_{\sssize 1.0}=U^n_{\sssize 1.0}(u^1,\ldots,u^n,x,y),
 \endcases
&
&\cases
 u^1_{\sssize 0.1}=U^1_{\sssize 0.1}(u^1,\ldots,u^n,x,y),\\
 \vspace{0.1ex}
 \hdotsfor 1\\
 \vspace{0.5ex}
 u^n_{\sssize 0.1}=U^n_{\sssize 0.1}(u^1,\ldots,u^n,x,y).
 \endcases
 \hskip -2em
\tag13.1
\endxalignat
$$
Right hand sides of these equations are the functions of $n$
dependent variables $u^1,\ldots,u^n$, and two independent
variables $x$ and $y$. Differentiation of such function
with respect to $x$ and $y$ is given by the following
differential operators:
$$
\aligned
D_x&=\frac{\partial}{\partial x}+\sum^n_{i=1}
U^i_{\sssize 1.0}(u^1,\ldots,u^n,x,y)\,
\frac{\partial}{\partial u^i},\\
D_y&=\frac{\partial}{\partial y}+\sum^n_{i=1}
U^i_{\sssize 0.1}(u^1,\ldots,u^n,x,y)\,
\frac{\partial}{\partial u^i}.
\endaligned
\tag13.2
$$
From \thetag{13.1}, as a result of successive differentiations,
we get the series of formulas for the derivatives
$u^i_{\sssize p.q}$ of the arbitrary order:
$$
u^i_{\sssize p.q}=U^i_{\sssize p.q}(u^1,\ldots,u^n,x,y).
\tag13.3
$$
Right hand sides of the equations \thetag{13.3} are bound
with each other by the recurrent relationships of the form:
$$
\xalignat 2
&U^i_{\sssize p+1.q}=D_xU^i_{\sssize p.q},
&
&U^i_{\sssize p.q+1}=D_yU^i_{\sssize p.q}.
\tag13.4
\endxalignat
$$
Let's represent the relationships \thetag{13.4} by means
of diagrams:
$$
\CD
 @AAA @AAA @AAA\\
 U^i_{\sssize 2.0}@>>>U^i_{\sssize 2.1} @>>>U^i_{\sssize 2.2}@>>>
 \hdots\\
 @AAA @AAA @AAA \\
 U^i_{\sssize 1.0} @>>>U^i_{\sssize 1.1}@>>>U^i_{\sssize 1.2}@>>>
 \hdots\\
 @. @AAA @AAA\\
 @. U^i_{\sssize 0.1}@>>>U^i_{\sssize 0.2}@>>>
 \hdots\\
\endCD\hskip -2em
\tag13.5
$$
Vertical arrows in \thetag{13.5} represent the action of the
operator $D_x$, while horizontal ones represent the action
of the operator $D_y$ from \thetag{13.2}. Recurrent relationships
\thetag{13.4} shown in the diagrams \thetag{13.5} are redundant.
They define uniquely only the side vertices in the diagrams
\thetag{13.5}, which have only one incoming arrow. Other
vertices with two incoming arrows give rise to the compatibility
conditions. Thus in the vertex $U^i_{\sssize p+1.q+1}$ we have
the equation
$$
D_xU^i_{\sssize p.q+1}=D_yU^i_{\sssize p+1.q}.
\tag13.6
$$
\proclaim{Lemma 13.1} The set of compatibility conditions
\thetag{13.6} written for the vertices $U^i_{\sssize 1.1}$
with $i=1,\ldots,n$ is equivalent to the permutability of
the operators $D_x$ and $D_y$ defined by \thetag{13.2}.
\endproclaim
\demo{Proof} Let's calculate the commutator $[D_x,\,D_y]$
of the operators $D_x$ and $D_y$ in explicit form. This
is the following differential operator of the first order:
$$
[D_x,\,D_y]=\sum^n_{i=1}\left(D_xU^i_{\sssize 0.1}-
D_yU^i_{\sssize 1.0}\right)\frac{\partial}{\partial u^i}.
\tag13.7
$$
The relationships \thetag{13.6} for the vertices
$U^i_{\sssize 1.1}$ are written as: $D_xU^i_{\sssize 0.1}=
D_yU^i_{\sssize 1.0}$. It is easy to see that they hold
for all $i=1,\ldots,n$ if and only if the commutator
\thetag{13.7} is equal to zero. Lemma is proved.
\qed\enddemo
\proclaim{Lemma 13.2} If $[D_x,D_y]=0$, then the
compatibility conditions \thetag{13.6} are fulfilled
for all vertices $U^i_{\sssize p+1.q+1}$ in the
diagrams \thetag{13.5}.
\endproclaim
\demo{Proof} For the vertex $U^i_{\sssize p.q}$ let's consider
the number $r=p+q$. We shall call it {\it the order} of this
vertex. We shall prove lemma by induction in $r$. The result
of lemma~13.1 forms the base for such induction for $r=2$.\par
    Suppose that lemma is proved for all vertices of the order
not greater than $p+q+1$. The vertex $U^i_{\sssize p+1.q+1}$
of the order $r=p+q+2$ is in the following rectangular part
of the diagram \thetag{13.5}:
$$
\CD
 U_{\sssize p+1.q} @>>> U_{\sssize p+1.q+1}\\
 @AAA @AAA\\
 U_{\sssize p.q} @>>> U_{\sssize p.q+1}\\
\endCD
$$
Due to inductive hypothesis, the compatibility relationship
\thetag{13.6} for $U^i_{\sssize p+1.q+1}$ can be rewritten
as:
$$
D_xU^i_{\sssize p.q+1}-D_yU^i_{\sssize p+1.q}=
D_xDyU^i_{\sssize p.q}-D_yD_xU^i_{\sssize p.q}=
[D_x,\,D_y]U^i_{\sssize p.q}=0.
$$
Now its clear that this relationship can be derived from
$[D_x,\,D_y]=0$. Lemma is proved.
\qed\enddemo
\proclaim{Theorem 13.1} Complete system of Pfaff equations
\thetag{13.1} is compatible if and only if the corresponding
operators \thetag{13.2} are commutating.
\endproclaim
     Theorem~13.1 is the consequence of lemmas~13.1 and 13.2.
It gives us an effective tool for checking the compatibility
of complete Pfaff equations.
\proclaim{Theorem 13.2} Each complete compatible system of
Pfaff equations \thetag{13.1} is locally solvable.
\endproclaim
\demo{Proof} Suppose that the condition $[D_x,\,D_y]=0$ for
the system of equations \thetag{13.1} is fulfilled. In order
to prove the local solvability for these equations,
let's state the Cauchy problem with the following initial
data for them:
$$
u^i\,{\vrule height 8pt depth 13pt}
 \lower 7pt
 \hbox{\vtop{\baselineskip=0pt
             \lineskip=2pt
             \lineskiplimit=0pt
             \hbox{$\,\ssize x=0$}
             \hbox{$\,\ssize y=0$}}}
             =\alpha^i_0.
\tag13.8
$$
For to solve this Cauchy problem \thetag{13.8}, note that
each system of equations in \thetag{13.1} can be treated as
a system of ordinary differential equations in $x$ and in
$y$ separately. For the second system in \thetag{13.1} we
consider the standard Cauchy problem
$$
u^i\,{\vrule height 8pt depth 8pt}
 \lower 7pt\hbox{$\,\ssize y=0$}=\alpha^i(x).
\tag13.9
$$
For the initial value functions $\alpha^i$ in \thetag{13.9}
from the first system \thetag{13.1} we get
$$
\cases
 (\alpha^1)'_x=U^1_{\sssize 0.1}(\alpha^1,\ldots,
 \alpha^n,x,0),
 \vspace{0.1ex}
 \hdotsfor 1\\
 \vspace{0.5ex}
 (\alpha^n)'_x=U^n_{\sssize 0.1}(\alpha^1,\ldots.
 \alpha^n,x,0),\\
\endcases
\tag13.10
$$
It is easy to see that \thetag{13.10} is a system of
ordinary differential equations in $x$. For this system,
from \thetag{13.8}, we derive:
$$
\alpha^i\,{\vrule height 8pt depth 8pt}
 \lower 7pt\hbox{$\,\ssize x=0$}=\alpha^i_0.
\tag13.11
$$
It is clear that \thetag{13.11} is a standard Cauchy
problem for \thetag{13.10}.\par
     As a result of successive solution of two standard
Cauchy problems \thetag{13.11} and \thetag{13.9}, we get
the set of functions $u^1(x,y),\ldots,u^n(x,y)$ which
satisfies the second system of Pfaff equations in
\thetag{13.1}. \pagebreak Now we are only to prove that these
functions satisfy the first system \thetag{13.1} too.
In order to do this, let's substitute $u^1(x,y),\ldots,
u^n(x,y)$ into the first system
$$
\matrix
 u^1_{\sssize 1.0}=U^1_{\sssize 1.0}(u^1,\ldots,u^n,x,y)+
 \delta^1,\\
 \vspace{0.1ex}
 \hdotsfor 1\\
 \vspace{0.6ex}
 u^n_{\sssize 1.0}=U^n_{\sssize 1.0}(u^1,\ldots,u^n,x,y)+
 \delta^n,
\endmatrix
\tag13.12
$$
and calculate the residuals $\delta^i(x,y)$. Then we differentiate
the relationships \thetag{13.12} with respect to $y$. For the
residuals $\delta^i(x,y)$ this gives:
$$
\delta^i_{\sssize 0.1}=u^i_{\sssize 1.1}-
\left(\frac{\partial U^i_{\sssize 1.0}}{\partial y}+
\shave{\sum^n_{j=1}}u^j_{\sssize 0.1}\,\frac{\partial
U^i_{\sssize 1.0}}{\partial u^j}\right)=
D_xU^i_{\sssize 0.1}-D_yU^i_{\sssize 1.0}-
\sum^n_{j=1}\frac{\partial U^i_{\sssize 1.0}}
{\partial u^j}\,\delta^j.
$$
But $D_xU^i_{\sssize 0.1}=D_yU^i_{\sssize 1.0}$, which is
the consequence of the compatibility of the equations
\thetag{13.1}. Therefore the residuals $\delta^i$ satisfy
the system of homogeneous differential equations in $y$
of the following form:
$$
\delta^i_{\sssize 0.1}=-\sum^n_{j=1}\frac{\partial
U^i_{\sssize 1.0}}{\partial u^j}\,\delta^j.
\tag13.13
$$
From \thetag{13.10} and \thetag{13.12} we derive zero initial
data for the equations \thetag{13.13}:
$$
\delta^i\,{\vrule height 8pt depth 8pt}
 \lower 7pt\hbox{$\,\ssize y=0$}=0.
\tag13.14
$$
From \thetag{13.13} and \thetag{13.14} we obtain the identical
vanishing of the residuals $\delta^i(x,y)$. Theorem is proved.
\qed\enddemo
    From the above proof of the theorem~13.2 we see that
the solution of the Cauchy problem \thetag{13.8} for the
compatible system of Pfaff equations is unique. Therefore
the general solution of such system of equations is
$n$-parametric family of functions parameterized by
constants $a^i_0$ in \thetag{13.8}.
\head
14. Appendix B. Pfaff equations with restrictions.
\endhead
     Let's consider the Pfaff equations \thetag{13.1}.
Right hand sides of these equations contain the independent
variables $x$ and $y$ in explicit form. Such Pfaff equations
are called {\it nonholonomic}. But each nonholonomic system
of Pfaff equations can be transformed to the {\it holonomic}
one by increasing its dimension $n\to n+2$. Let's add two
dependent variables $u^{n+1}=x$ and $u^{n+2}=y$ and define
four new functions
$$
\xalignat 2
&U^{n+1}_{\sssize 1.0}(u^1,\ldots,u^{n+2})=1,
&&U^{n+1}_{\sssize 0.1}(u^1,\ldots,u^{n+2})=0,\\
\vspace{-1ex}
&&&\tag14.1\\
\vspace{-1ex}
&U^{n+2}_{\sssize 1.0}(u^1,\ldots,u^{n+2})=0,
&&U^{n+2}_{\sssize 1.0}(u^1,\ldots,u^{n+2})=1.
\endxalignat
$$
These functions \thetag{14.1} let us to add two new equations
to each system of Pfaff equations in \thetag{13.1}. On doing
this, we get complete holonomic system of Pfaff equations of
the dimension $n+2$. This operation is called {\it holonomic
expansion} of the equations \thetag{13.1}. It is easy to check
that the operation of holonomic expansion preserves the
compatibility of Pfaff systems.\par
     Operations, that diminish the dimension of Pfaff equations
are called {\it restrictions}. Regular way of doing this is
connected with adding some restricting equations to the system.
Let's consider the complete holonomic system of Pfaff equations:
$$
\xalignat 2
&\cases
 u^1_{\sssize 1.0}=U^1_{\sssize 1.0}(u^1,\ldots,u^n),\\
 \vspace{0.1ex}
 \hdotsfor 1\\
 \vspace{0.5ex}
 u^n_{\sssize 1.0}=U^n_{\sssize 1.0}(u^1,\ldots,u^n),
 \endcases
&
&\cases
 u^1_{\sssize 0.1}=U^1_{\sssize 0.1}(u^1,\ldots,u^n),\\
 \vspace{0.1ex}
 \hdotsfor 1\\
 \vspace{0.5ex}
 u^n_{\sssize 0.1}=U^n_{\sssize 0.1}(u^1,\ldots,u^n).
 \endcases
 \hskip -2em
\tag14.2
\endxalignat
$$
Then consider the following system of functional equations:
$$
\cases
 I^1(u^1,\ldots,u^n)=0,\\
 \vspace{0.1ex}
 \hdotsfor 1\\
 \vspace{0.5ex}
 I^k(u^1,\ldots,u^n)=0.
 \endcases
\tag14.3
$$
This system of equations \thetag{14.3} is assumed to be
{\it regular}. Such system defines $(n-k)$-dimensional
submanifold $M$ in the space $\Bbb R^n$ of variables
$u^1,\ldots,u^n$. Each solution of Pfaff equations
\thetag{14.2} defines two-parametric subset in the same
space $\Bbb R^n$. If $S\subset M$, then we have the
following relationships
$$
\xalignat 2
&\sum^n_{q=1}\frac{\partial I^k}{\partial u^q}\,
U^q_{\sssize 1.0}=0,
&&\sum^n_{q=1}\frac{\partial I^k}{\partial u^q}\,
U^q_{\sssize 0.1}=0.
\tag14.4
\endxalignat
$$
The relationships \thetag{14.4} have the same structure as
the equations \thetag{14.3}. We say that the restrictions
\thetag{14.3} are {\it consistent} with Pfaff equations
\thetag{14.2}, if the relationships \thetag{14.4} are
functional consequences of the equations \thetag{14.3},
i\. e\. if they hold identically on the manifold $M$.
\definition{Definition 14.1} Differential Pfaff equations
\thetag{14.2} equipped with consistent functional equations
\thetag{14.3} are called {\it restricted} Pfaff equations.
\enddefinition
    If the condition of consistence is not fulfilled, we
can add the equations \thetag{14.4} to the system \thetag{14.3}.
Then we extract the maximal functionally independent subsystem
of equations in this enlarged system. It is clear that such
subsystem will be consistent with \thetag{14.2}. Therefore
we can always consider only the consistent systems of
restrictions \thetag{14.3}.\par
     Now let's remember the operators $D_x$ and $D_y$ from
\thetag{13.2}. Here they can be treated as the vector fields
in $\Bbb R^n$:
$$
\aligned
D_x&=\sum^n_{i=1}
U^i_{\sssize 1.0}(u^1,\ldots,u^n)\,
\frac{\partial}{\partial u^i},\\
D_y&=\sum^n_{i=1}
U^i_{\sssize 0.1}(u^1,\ldots,u^n)\,
\frac{\partial}{\partial u^i}.
\endaligned
\tag14.5
$$
The consistence of \thetag{14.3} and \thetag{14.2} means
that vector fields \thetag{14.5} are tangent to the
submanifold $M$. It is known that tangent fields can be
restricted to submanifold. Let $v^1,\ldots,v^{n-k}$ be
the system of local coordinates on $M$. Then the restrictions
of the vector field \thetag{14.5} on $M$ can be represented
as:
$$
\aligned
\tilde D_x&=\sum^{n-k}_{i=1}
V^i_{\sssize 1.0}(v^1,\ldots,v^{n-k})\,
\frac{\partial}{\partial v^i},\\
\tilde D_y&=\sum^{n-k}_{i=1}
V^i_{\sssize 0.1}(v^1,\ldots,v^{n-k})\,
\frac{\partial}{\partial v^i}.
\endaligned
\tag14.6
$$
These vector fields \thetag{14.6} can define some $(n-k)$-dimensional
system of Pfaff equations on $M$
$$
\xalignat 2
&\quad\cases
 v^1_{\sssize 1.0}=V^1_{\sssize 1.0}(v^1,\ldots,v^{n-k}),\\
 \vspace{0.1ex}
 \hdotsfor 1\\
 \vspace{0.5ex}
 v^{n-k}_{\sssize 1.0}=V^{n-k}_{\sssize 1.0}(v^1,\ldots,u^{n-k}),
 \endcases
&
&\cases
 v^1_{\sssize 0.1}=V^1_{\sssize 0.1}(v^1,\ldots,v^{n-k}),\\
 \vspace{0.1ex}
 \hdotsfor 1\\
 \vspace{0.5ex}
 v^{n-k}_{\sssize 0.1}=V^{n-k}_{\sssize 0.1}(v^1,\ldots,v^{n-k}).
 \endcases
 \hskip -2em
\tag14.7
\endxalignat
$$
It is obvious that each solution the equations \thetag{14.7}
defines some solution for \thetag{14.2}, which is the solution
for \thetag{14.3} too. And conversely, each solution of the
restricted Pfaff equations \thetag{14.2} and \thetag{14.3}
gives some solution for \thetag{14.7}. Therefore the system
of Pfaff equations \thetag{14.7} is called {\it the restriction}
of the equations \thetag{14.2} due to \thetag{14.3}.
\par
    The following two facts are well known in differential
geometry: commutator of two vector fields tangent to the
submanifold $M$ is also tangent to $M$; the restriction of
commutator for two tangent vector fields coincides with the
commutator of their restrictions. These facts make natural
the following definition.
\definition{Definition 14.2} Complete system of restricted
Pfaff equations \thetag{14.2} with restrictions \thetag{14.3}
is called {\it compatible}, if the commutator of corresponding
differential operators \thetag{14.5} vanishes at any point of
submanifold $M$ defined by \thetag{14.3}.
\enddefinition
    It is clear that the compatibility of the Pfaff equations
\thetag{14.2} with restrictions \thetag{14.3} in the sense of
definition~14.2 is equivalent to the compatibility of its
restriction \thetag{14.7} in the sense of theorem~13.1.
Therefore from the results of previous section we derive
the following theorem.
\proclaim{Theorem 14.1} Each complete compatible system of
Pfaff equations \thetag{14.2} with restrictions \thetag{14.3}
is locally solvable and in some neighborhood of any point
$(x,y)$ it possess $(n-k)$-parametric set of solutions.
\endproclaim
\head
15. Acknowledgments.
\endhead
     Author is grateful to Professors E.G.~Neufeld, V.V.~Sokolov,
A.V.~Bocharov, V.E.~Adler, N.~Kamran, V.S.~Dryuma, A.B.~Sukhov
and M.V.~Pavlov for the information, for worth advises, and for
help in finding many references below.\par

\enddocument
\end